\documentclass[12pt,draftcls,onecolumn]{IEEEtran}
\usepackage{graphicx}
\usepackage{subfigure}
\usepackage[cmex10]{amsmath}
\usepackage{amssymb}
\usepackage{bm}
\usepackage{algorithm}
\usepackage{algorithmic}
\usepackage{cite}
\usepackage{setspace}
\usepackage{stfloats}
\usepackage{multirow}
\usepackage{mathrsfs}
\usepackage{color}
\usepackage{verbatim}
\usepackage{extpfeil}
\usepackage{hyperref}
\usepackage{caption}
\begin{document}
\makeatletter
\newcommand{\rmnum}[1]{\romannumeral #1}
\newcommand{\Rmnum}[1]{\expandafter \@slowromancap \romannumeral #1@}
\makeatother

\title{Joint Angle Estimation Error Analysis and 3D Positioning Algorithm Design for mmWave Positioning System}

\author{Tuo Wu, Cunhua Pan, Yijin Pan, Sheng Hong, Hong Ren, Maged Elkashlan, Feng Shu and Jiangzhou Wang, \emph{Fellow, IEEE}

\thanks{(Corresponding author: Cunhua Pan).
	
T. Wu and M. Elkashlan are with the School of Electronic Engineering and Computer Science at Queen
Mary University of London, London E1 4NS, U.K. (Email:\{tuo.wu, maged.elkashlan\}@qmul.ac.uk).  C. Pan, Y. Pan and H. Ren are with the National Mobile Communications Research Laboratory, Southeast University, Nanjing 210096, China. (Email: \{cpan,panyj,hren\}@seu.edu.cn). S. Hong is with Information Engineering School of Nanchang University, Nanchang 330031, China. (Email: shenghong@ncu.edu.cn).
F.  Shu is with the School of Electronic and Optical Engineering, Nanjing
University of Science and Technology, Nanjing 210094, China, and also
with the School of Information and Communication Engineering, Hainan
University, Haikou 570228, China.(E-mail: shufeng0101@163.com). J.
Wang is with the School of Engineering, University of Kent, UK. (e-mail: J.Z.Wang@kent.ac.uk).
}

}

\markboth{}
{}

\maketitle
\begin{abstract}
   In this paper, we propose a comprehensive framework to jointly analyze the angle estimation error and design the three-dimensional (3D) positioning algorithm for a millimeter wave (mmWave) positioning system. First, we estimate the angles of arrival (AoAs) at the anchors by applying the two-dimensional discrete Fourier transform (2D-DFT) algorithm.  Based on the property of the 2D-DFT algorithm,  the angle estimation error is analyzed in terms of probability density functions (PDF). The analysis shows that the derived angle estimation error is non-Gaussian, which is different from the existing work.  Second, the intricate expression of the PDF of the AoA estimation error is simplified by employing the first-order linear approximation of triangle functions. Then, we derive a complex expression for the variance based on the derived PDF. Specifically, for the azimuth estimation error, the variance is separately integrated according to the different non-zero intervals of the PDF. Finally,  we apply the weighted least square (WLS) algorithm to estimate the 3D position of the MU by using the estimated AoAs and the obtained non-Gaussian variance. Extensive simulation results confirm that the derived angle estimation error is non-Gaussian, and also demonstrate the superiority of the proposed  framework.
\end{abstract}
\IEEEpeerreviewmaketitle

\begin{comment}More specifically, we first derive the PDF by using the geometric relationship between the AoAs and their triangle functions.We  consider the non-Gaussian angle estimation error based on the practical angle estimation method, e.g, two-dimensional discrete Fourier transform (2D-DFT) algorithm. propose a  derivation algorithm to model and
	 enabled by a combination of wide band signals (up to 400 MHz), higher carrier frequency (e.g., around 28GHz), lower latency and multiple antennas
\end{comment}
\section{Introduction}
As an important requirement for the sixth generation (6G) wireless networks, high-accuracy positioning service is of great importance in a wide range of applications \cite{ZhouY}, such as  automated driving vehicles \cite{Ding1}, smart factory \cite{Ren4} and virtual reality \cite{Dengy}.
For example, it is predicted that the 2020s will be the first decade for automated driving vehicles  with the positioning accuracy at decimeter level \cite{AV}. However, the positioning accuracy of the prevalent global positioning system (GPS) is about 5 meters even in ideal conditions \cite{Sari}, which cannot meet the stringent requirement on positioning accuracy for these thriving applications \cite{Mendrzik}. Therefore, network-based positioning systems are emerging as a promising alternative to GPS in 6G networks \cite{Liu}.

In current wireless positioning networks, the millimeter wave (mmWave) technique \cite{Pan1} can provide an extremely highly-accurate estimation of channel parameters, such as the channel gain, time delays, and angle of arrival (AoA).  These parameters can be used to estimate the position of the mobile user (MU). Therefore, it is of much interest to explore positioning algorithms for the mmWave positioning systems. Typically, positioning algorithms work through the following two steps. First,  channel parameters can be acquired through some estimation methods \cite{Dardari, QianTD, Gao1}.  Second,  positioning algorithms can be designed accordingly. Specifically,  expressions of the complex non-linear geometric relationships between the channel parameters and the position coordinates are first derived \cite{Lee3}. Then, the estimation error of the channel parameters and the non-linear geometric relationships are jointly utilized to derive the non-linear equations  \cite{Lee2}. Finally,  the position of the MU is determined by solving the equations using iterative or non-iterative algorithms. From the above-mentioned positioning steps,  the estimation error of the channel parameters can directly determine the positioning accuracy. In general,  different methods for estimating different channel parameters lead to different types of parameter estimation errors \cite{Daill1}.

To date, mmWave positioning systems have attracted extensive research attention \cite{Lin1,ZBp,JHe,Alouini, Kanh,Jin11}.  To reduce the high computational complexity due to a large number of antennas, \cite{Lin1}  proposed a novel channel compression method for the mmWave positioning systems.  In \cite{ZBp}, the successive localization and beamforming scheme was proposed to estimate the long-term MU location and the instantaneous channel state for the mmWave multiple-input multiple-output (MIMO) communications. To provide an analytical performance validation, \cite{JHe} studied the theoretical performance bounds (i.e., Cram$\rm{\acute{e}}$r-Rao lower bounds (CRLB)) for positioning, and evaluated the impact of the number of reflecting elements and the phase shifts of reconfigurable intelligent surface (RIS) on the positioning estimation accuracy.  Alouini \emph{et al.} derived the CRLB for assessing the performance of synchronous and asynchronous signaling schemes and proposed an optimal closed-form expression of the reflecting phase shifts of the RIS for joint communication and localization \cite{Alouini}.  The authors of \cite{Kanh} demonstrated that accurate estimates of the position of an unknown node can be determined using estimates of time of arrival (ToA), and AoA, as well as data fusion or machine learning. \cite{Jin11} considered the channel estimation problem and the channel-based wireless applications in MIMO orthogonal frequency division multiplexing (OFDM) systems assisted by RISs.

However, the above contributions assumed that  the estimation error of the channel parameters follows the Gaussian distribution, which is inconsistent with practical scenarios. In practice,  the distribution of these parameters depends on the practical estimation methods (e.g., 2D-DFT algorithm \cite{Zhou1}), which may not follow the Gaussian distribution. Therefore, existing positioning algorithms and performance analysis may not be applicable when considering practical channel parameter estimation methods. Hence, it is necessary to model the estimation error of the estimated channel parameters, so that the position of the MU can be estimated based on the estimation error.

In this paper, we aim to propose a complete framework to  jointly analyze the angle estimation and design the three-dimensional (3D) positioning algorithm for the mmWave positioning system. Our main contributions are summarized as follows:
\begin{itemize}
\item[1)] We propose a comprehensive framework to jointly  analyze the angle estimation and design the 3D positioning algorithm for the mmWave positioning system.  Specifically,  we first estimate the AoAs at the anchors by applying the 2D-DFT algorithm. Then, we derive the closed-form expression of the PDF of the estimation error. We further derive the variance of the estimation error based the PDF. Finally, by using the estimated AoAs and the derived variance, we apply the weighted least square (WLS) algorithm to estimate the 3D position of the MU.

\item[2)] The estimated AoAs are first derived in closed-form based on the 2D-DFT method.  Based on the expression of estimation, we find that the angle estimation  error at the anchors depends on the search grid, the panel size of the anchors, and the number of antennas of the anchors.  Moreover, due to the property of 2D-DFT,  the angle estimation error follows the uniform distribution.

\item[3)] According to the uniform distribution of the estimation error of azimuth and elevation, the angle estimation error is characterized in terms of the PDF, which is non-Gaussian.  To be specific, we first derive the PDF by using the geometric relationship between the AoAs and their triangle functions. Then, we simplify the complex geometric expression of the PDF by employing the first-order linear approximation of the triangle function. For the azimuth angle estimation, we provide an algorithm to derive and approximate the PDF expression of its estimation error.

\item[4)]Based on the PDF of estimation error, we theoretically derive the variance of the angle estimation error. Since the PDF of the azimuth estimation error has three different non-zero intervals, we separately calculate the integral in the different intervals according to the variance calculation formula.

\item[5)]Simulation results verify the accuracy of  the derived results and demonstrate the superiority of the proposed framework. We observe that the variance decreases with the number of elements, which means that increasing the number of anchor antennas improves the estimation accuracy.
\end{itemize}

The remainder of the paper is organized as follows. The system model for the mmWave positioning system is described in Section \ref{System Model}. The details of the whole framework are given in Section \ref{aa}. The procedures of estimating angles are given in Section \ref{angle estimation algorithm}. Section \ref{PDF} derives the PDF in more details. Section \ref{Variance} calculates the variance. The positioning algorithm is given in Section \ref{WLS}. Simulation results are given in Section \ref{result}. Section \ref{Con} concludes the work of this paper.

\begin{figure}[!ht]
	\centering
	{\includegraphics[width=4.5 in]{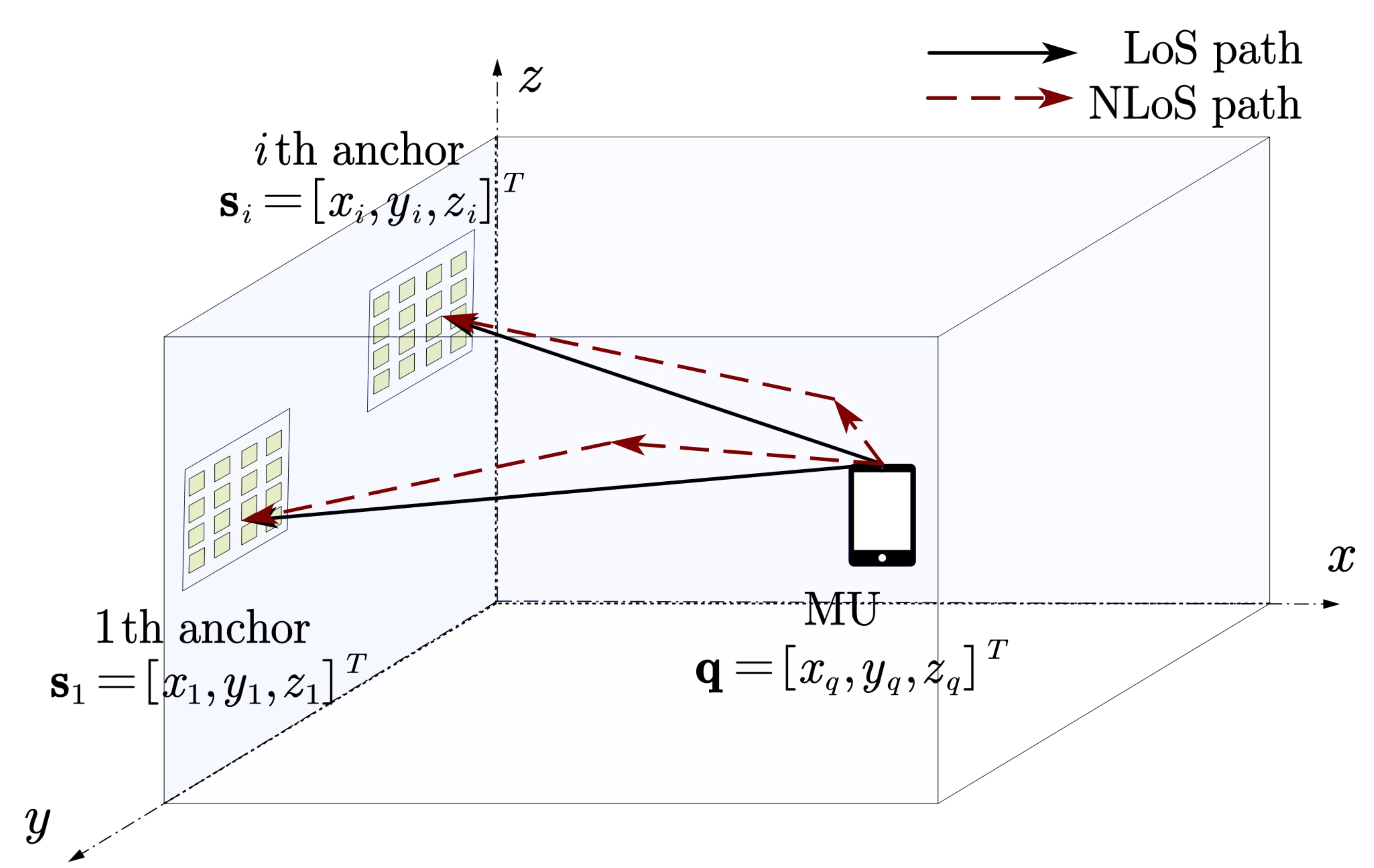}}
	\caption{\normalsize System model of considered mmWave positioning system.}\label{system model}
\end{figure}
\section{System Model} \label{System Model}

Consider a mmWave  time division duplex (TDD) 3D positioning system, where an MU sends pilot signals to the anchors
 to locate the MU. We assume that there are $I$ anchors, each of which is equipped with a uniform planar array (UPA) with $N_{y,z}=N_y\times N_z$ antennas, where $N_y$ and $N_z$ denote the numbers of antennas along the y-axis and z-axis, respectively. The MU is equipped with a single antenna.

As shown in Fig. \ref{system model}, the anchors are placed parallel to the y-o-z plane  with the center located at $\mathbf{s}_i=[x_i,y_i,z_i]^T$, $i\in[1,\cdots,I]$. The true location of the MU is $\mathbf{q}=[x_q,y_q,z_q]^T$. The estimated location of the MU is denoted as $\mathbf{\hat{q}}=[\hat{x}_q,\hat{y}_q,\hat{z}_q]^T$.
    Generally, once the anchors are deployed, the coordinate $\mathbf{s}_i$ are known and invariant. In order to locate the MU, we need to obtain the estimated $\mathbf{\hat{q}}$.
\begin{comment}
\begin{figure}[!t]
  \subfigure[Transmission link from the MU to the RIS.]{  \includegraphics[width=0.48\linewidth]{AoA_RIS}\label{AoA_RIS}
}\subfigure[Transmission link from the RIS to the anchor.]{  \includegraphics[width=0.48\linewidth]{AOD_RIS}\label{anchor-RIS}
}\caption{\normalsize Two main parts of the system model.}
\label{system model1}
\end{figure}
\end{comment}

Assuming that the number of propagation paths between the  MU and the $i$th anchor is $N_{i}$, the AoA of the $n_i$th path from the MU to the $i$th anchor can be decomposed into the elevation angle $0\leq\Theta_{n_i}\leq\pi$ in the vertical direction, and the azimuth angle $0\leq\Phi_{n_i}\leq\pi$ in the horizontal direction, respectively.  As a result, the array response vector at the $i$th anchor of the $n_i$th path can be expressed as
\begin{align}\label{1}
{\bf a}(\Theta_{n_i},\Phi_{n_i})={\bf a}_{e}(\Phi_{n_i})\otimes{\bf a}_{a}(\Theta_{n_i},\Phi_{n_i}),
\end{align}
where $\otimes$ denotes the Kronecker product. Moreover, we have
\begin{align}\label{2}
{\bf a}_{e}(\Phi_{n_i})=[1,e^{\frac{-j2\pi d_r\sin\Phi_{n_i}}{\lambda_c}},...,e^{\frac{-j2\pi(M_{x}-1) d_r\sin\Phi_{n_i}}{\lambda_c}}]^T,
\end{align}
and
\begin{align}\label{3}
{\bf a}_{a}(\Theta_{n_i},\Phi_{n_i})=[1,e^{\frac{-j2\pi d_r\cos\Theta_{n_i}\cos\Phi_{n_i}}{\lambda_c}},...,
e^{\frac{-j2\pi(M_{z}-1) d_r\cos\Theta_{n_i}\cos\Phi_{n_i}}{\lambda_c}}]^T,
\end{align}
where $d_r$ and $\lambda_c$ denote the distance between the antennas of the anchors and  the carrier wavelength, respectively. Then, the channel between the MU and the $i$th anchor, denoted as ${\bf h}_i$,  can be modeled as
\begin{align}\label{4}
{\bf h}_i =\sum^{N_i}_{n_i=1}\alpha_{n_i}{\bf a}(\Theta_{n_i},\Phi_{n_i})= \underbrace{\alpha_{L,i}{\bf a}(\Theta_{L,i},\Phi_{L,i})}_{LoS}+ \underbrace{\sum^{N_i}_{n_i=2}\alpha_{n_i}{\bf a}(\Theta_{n_i},\Phi_{n_i})}_{NLoS},
\end{align}
where  $\alpha_{n_i}$ denotes the complex channel gain of $n_i$th path. Moreover, $\alpha_{L,i}$, $\Theta_{L,i}$,and $\Phi_{L,i}$ denote the complex channel gain, the elevation AoA, and the azimuth AoA of the line-of-sight (LoS) path, respectively. As we can see from \eqref{4}, channel components of ${\bf h}_i$  can be categorized into two types, namely \emph{LoS} and \emph{NLoS}. LoS path component is the direct path between the anchors and the MU, non-line-of-sight (NLoS) path component consists of the paths between the anchors and the MU reflected by scatters, e.g., walls, human bodies, and etc. Moreover, according to \cite{basar2019wireless}, the complex channel gain of LoS is given by
\begin{align}\label{ap1}
\alpha_{L,i}=\frac{\lambda_c e^{-j2\pi d_{L,i}}}{4\pi d_{L,i}},
\end{align}
where $d_{L,i}$ is the distance between the $i$th anchor and the MU.

\section{Joint Angle Estimation Analysis and 3D Positioning Algorithm Design}\label{aa}
In the existing works concerning the positioning algorithm design,  for tractability,  the estimation error of channel parameters is assumed to be additive zero-mean complex Gaussian noise. However, in practice,  the distribution of these channel parameters depends on the practical estimation methods, which may not follow the Gaussian distribution. To investigate the impact of the practical estimation error of the channel parameters,  we propose to design a comprehensive framework to jointly analyze the angle estimation error and estimate the 3D position of the MU.

Based on the 2D-DFT angle estimation technique,  the angle estimation error analysis and 3D positioning algorithm design is investigated in this paper. First, we apply the 2D-DFT algorithm to estimate $( \Theta_{L,i},\Phi_{L,i} )$  \footnote{As NLoS path component usually varies fast and its weight to the channel is marginal, especially in the mmWave band, we are more interested in LoS path. Hence, we intend to estimate the path parameter $( \Theta_{L,i},\Phi_{L,i} )$ of the LoS component from the MU to the anchors, which can be used to derive the position of the MU.}. Then, based on the estimation of the AoAs, we first derive the  PDF of the angle estimation error, based on which the closed-form expression of the variance of the angle estimation error is derived. Finally, we apply the WLS algorithm to estimate the 3D position of the MU by using the estimated AoAs and the derived variance.

 The details of the proposed framework are summarized in Algorithm \ref{JAEAP}.  The descriptions of each step of the proposed framework will be introduced in the following sections.
\begin{algorithm}
	\caption{Joint Angle Estimation Error Analysis and 3D Positioning Framework}
	\label{JAEAP}
	\begin{algorithmic}[1]
        \STATE Estimate $( \Theta_{L,i},\Phi_{L,i} )$ by using the 2D-DFT estimation algorithm;
		\STATE Derive the PDF of the angle estimation error;
		\STATE Derive the closed-form expression of the variance of the angle estimation error;
        \STATE Estimate  the 3D position of the MU by using the variance of the angle estimation error.
	\end{algorithmic}
\end{algorithm}

\section{Angle Estimation} \label{angle estimation algorithm}
The first step of the proposed framework is to estimate $( \Theta_{L,i},\Phi_{L,i} )$. Hence, in this section, we apply the 2D-DFT algorithm \cite{Gao1} to estimate the AoA  at the anchors. For the sake of illustration, we consider the estimation errors only for the noise-free scenario similar to \cite{Gao1} and \cite{Yang}, the performance of which is roughly the same as the general scenarios with sufficiently high received signal to noise ratio (SNR).
\subsection{Initial Angle Estimation}
According to the expression of ${\bf a}(\Theta_{L,i},\Phi_{L,i})$ in \eqref{1}, ${\bf a}(\Theta_{L,i},\Phi_{L,i})$ can be further derived as
\begin{align}\label{av1}
{\bf a}(\Theta_{L,i},\Phi_{L,i})={\rm vec}\{{\bf A}(\Theta_{L,i},\Phi_{L,i})\}= {\rm vec}\{{\bf a}_e(\Phi_{L,i}){\bf a}_a(\Theta_{L,i},\Phi_{L,i})\}.
\end{align}
To estimate the AoA at the $i$th anchor, we define two normalized DFT matrices ${\bf F}_{N_y}$ and ${\bf F}_{N_z}$,  elements of which are written as $[{\bf F}_{N_y}]_{b_{i}b_{i}'} = e^{-j\frac{2\pi}{N_y}b_{i}b_{i}'}$ $(b_{i},b_{i}'=0,1,\cdots,N_y-1)$
  and $[{\bf F}_{N_z}]_{qq'} = e^{-j\frac{2\pi}{N_z}q_{i}q_{i}'}$ $(q_{i},q_{i}'=0,1,\cdots,N_z-1)$, respectively.
 Meanwhile, let us define $u_{L,i}=\frac{2\pi d_r\cos\Theta_{L,i}\cos\Phi_{L,i}}{\lambda_c}$ and $ v_{L,i}=\frac{ 2\pi d_r\sin\Phi_{L,i}}{\lambda_c}$. Then, we  define the normalized 2D-DFT of the matrix ${\bf A}(\Theta_{L,i},\Phi_{L,i})$ in \eqref{av1} as
  ${\bf A}_{DFT}(\Theta_{L,i},\Phi_{L,i})={\bf F}_{N_y}{\bf A}(\Theta_{L,i},\Phi_{L,i}){\bf F}_{N_z}$, whose $(b_{i},q_{i})$th element is calculated as
\begin{align}
[{\bf A}_{DFT}(\Theta_{L,i},\Phi_{L,i})]_{b_{i}q_{i}}&=\sum^{N_y-1}_{n_y=0}\sum^{N_z-1}_{n_z=0}[{\bf A}(\Theta_{L,i},\Phi_{L,i})]_{b_{i}q_{i}}e^{-j2\pi(\frac{b_{i}n_y}{N_y}
+\frac{q_{i}n_z}{N_z})}\nonumber\\
&=e^{j\frac{N_y-1}{2}(u_{L,i}-\frac{2\pi b_{i}}{N_y})}e^{j\frac{N_z-1}{2}(v_{L,i}-\frac{2\pi q_{i}}{N_z})}\nonumber\\
&\quad\times \frac{\sin(\pi b_{i}-\frac{N_yu_{L,i}}{2})}{\sin((\pi b_{i}- \frac{N_yu_{L,i}}{2})/N_y)} \cdot
\frac{\sin(\pi q_{i}-\frac{N_zv_{L,i}}{2})}{\sin((\pi q_{i}-\frac{N_zv_{L,i}}{2})/N_z)}.
\end{align}
\begin{comment}
Firstly, we define two normalized DFT matrices ${\bf F}_{N_y}$ and ${\bf F}_{N_z}$,  elements of which are written as $[{\bf F}_{N_y}]_{pp'} = e^{-j\frac{2\pi}{N_y}pp'}$ $(p,p'=0,1,\cdots,N_y-1)$
  and $[{\bf F}_{N_z}]_{qq'} = e^{-j\frac{2\pi}{N_z}qq'}$ $(q,q'=0,1,\cdots,N_z-1)$, respectively.
 Meanwhile, we define the normalized 2D-DFT of the matrix $\textbf{a}_{L,i}$ in \eqref{1ee2} as
  $\textbf{a}_{L,i_{DFT}}={\bf F}_{N_y}\textbf{a}_{L,i}{\bf F}_{N_z}$, whose $(p,q)$th element is calculated as
\begin{align}
[\textbf{a}_{L,i_{DFT}}]_{pq}&=\sum^{N_y-1}_{n_y=0}\sum^{N_z-1}_{n_z=0}[\textbf{a}_{L,i}]_{pq}e^{-j2\pi(\frac{pn_y}{N_y}
+\frac{qn_z}{N_z})}\nonumber\\
&=e^{j\frac{N_y-1}{2}(u_{L,i}-\frac{2\pi p}{N_y})}e^{j\frac{N_z-1}{2}(v_{L,i}-\frac{2\pi q}{N_z})}\nonumber\\
&\quad\times \frac{\sin(\pi p-\frac{N_yu_{L,i}}{2})}{\sin((\pi p- \frac{N_yu_{L,i}}{2})/N_y)} \cdot
\frac{\sin(\pi q-\frac{N_zv_{L,i}}{2})}{\sin((\pi q-\frac{N_zv_{L,i}}{2})/N_z)}.
\end{align}
\end{comment}
When the number of reflecting elements becomes infinite,
i.e., $N_y\rightarrow\infty$, $N_z\rightarrow\infty$, there always exist some integers
$b_{n_i}=\frac{N_yu_{L,i}}{2\pi}$, $q_{n_i}=\frac{N_zv_{L,i}}{2\pi}$ such that $[{\bf A}_{DFT}(\Theta_{L,i},\Phi_{L,i})]_{b_{n_i}q_{n_i}}=1$,
while the other elements are all zero. Therefore,
all power is concentrated on the $(b_{n_i},q_{n_i})$th element and ${\bf A}_{DFT}(\Theta_{L,i},\Phi_{L,i})$ is a sparse matrix.
However, the anchor size could not be infinitely large, thus $\frac{N_yu_{L,i}}{2\pi}$ and $\frac{N_zv_{L,i}}{2\pi}$ may not
be integers in general, which leads to the channel power leakage from the $(b_{n_i},q_{n_i})$th element to its adjacent element.
However, ${\bf A}_{DFT}(\Theta_{L,i},\Phi_{L,i})$ can still be approximated as a sparse matrix with the most power concentrated around the $(b_{n_i},q_{n_i})$th element.
Therefore, the peak power position of ${\bf A}_{DFT}(\Theta_{L,i},\Phi_{L,i})$ is still useful for estimating the AoAs at the achor.
Then, the initial estimation is derived  as follows:
 \begin{comment}
\begin{align}\label{1e12}
\cos\hat{\theta}_{L,i}\cos\hat{\phi}_{L,i}=\frac{\lambda p_n}{N_y d_1 },\nonumber\\
\sin\hat{\phi}_{L,i}=\frac{\lambda q_n}{N_z d_1 }.
\end{align}
Base on the above discussion, we can derive the 2D-DFT of the estimated matrix $\hat\textbf{a}_{L,i}$, with its $(p,q)$th element being
\begin{align}\label{1e13}
[\hat\textbf{a}_{L,i}]_{pq}=[\textbf{a}_{L,i}]_{pq}+[{\bf N}_{L,i}]_{pq},
\end{align}
where $[{\bf N}_{L,i}]\sim\mathcal{CN}(0,\delta^2{\bf I})$,  and we can express the initial AoA estimates as
\end{comment}
\begin{align}\label{1e14}
\cos\hat{\Theta}^{ini}_{L,i}\cos\hat{\Phi}^{ini}_{L,i}=\frac{\lambda_c b_{n_i}}{N_y d_r },\nonumber\\
\sin\hat{\Phi}^{ini}_{L,i}=\frac{\lambda_c q_{n_i}}{N_z d_r },
\end{align}
where $\hat{\Theta}^{ini}_{L,i}$ and $\hat{\Phi}^{ini}_{L,i}$ denote the initial estimated angles at the anchor.
\subsection{Fine Angle Estimation}
The resolution of the estimated $\sin\hat{\Phi}^{ini}_{L,i}$ and $\cos\hat{\Theta}^{ini}_{L,i}\cos\hat{\Phi}^{ini}_{L,i}$
 is limited by the half of the DFT interval. In order to improve the estimation accuracy, angle rotation is provided to solve the mismatch issue in this subsection\cite{Zhou1}.

Let us define the angle rotation matrix of ${\bf A}(\Theta_{L,i},\Phi_{L,i})$ as ${\bf A}^{ro}(\Theta_{L,i},\Phi_{L,i})$, expressed as
\begin{align}\label{1e15}
{\bf A}^{ro}(\Theta_{L,i},\Phi_{L,i}) = {\bf U}_{N_y}(\Tilde{\varpi}_{1,i}){\bf A}(\Theta_{L,i},\Phi_{L,i}){\bf U}_{N_z}(\Tilde{\varpi}_{2,i}),
  \end{align}
where the diagonal matrices ${\bf U}_{N_y}(\Tilde{\varpi}_{1,i})$ and ${\bf U}_{N_z}(\Tilde{\varpi}_{2,i})$ are given by
\begin{align}\label{1e16}
{\bf U}_{N_y}(\Tilde{\varpi}_{1,i})={\rm Diag}\{1,e^{j\Tilde{\varpi}_{1,i}},...,e^{j(N_y-1)\Tilde{\varpi}_{1,i}}\},\nonumber\\
{\bf U}_{N_z}(\Tilde{\varpi}_{2,i})={\rm Diag}\{1,e^{j\Tilde{\varpi}_{2,i}},...,e^{j(N_z-1)\Tilde{\varpi}_{2,i}}\},
  \end{align}
with $\Tilde{\varpi}_{1,i}\in[-\pi/N_y,\pi/N_y]$ and $\Tilde{\varpi}_{2,i}\in[-\pi/N_z,\pi/N_z]$ being the angle rotation parameters.
By using the angle rotation operation, the $(b_i,q_i)$th element of the 2D-DFT of the rotated matrix ${\bf A}^{ro}_{DFT}(\Theta_{L,i},\Phi_{L,i})$ is calculated as
\begin{align}
[{\bf A}^{ro}_{DFT}(\Theta_{L,i},\Phi_{L,i})]_{b_iq_i}&=\sum^{N_y-1}_{n_y=0}\sum^{N_z-1}_{n_z=0}[{\bf A}(\Theta_{L,i},\Phi_{L,i})]_{b_iq_i}
e^{-j2\pi(\frac{b_in_y}{N_y}+\frac{q_in_z}{N_z})}e^{j2\pi(\frac{\Tilde{\varpi}_{1,i}n_y}{2\pi}
+\frac{\Tilde{\varpi}_{2,i}n_z}{2\pi})}\nonumber\\
&=e^{j\frac{N_y-1}{2}(u_{L,i}+\Tilde{\varpi}_{1,i}-\frac{2\pi b_i}{N_y})}
e^{j\frac{N_z-1}{2}(v_{L,i}+\Tilde{\varpi}_{2,i}-\frac{2\pi q_i}{N_z})}\nonumber\\
&\quad\times \frac{\sin(\pi b_i-\frac{N_yu_{L,i}}{2}-\frac{N_y\Tilde{\varpi}_{1,i}}{2})}{\sin((\pi b_i-\frac{N_yu_{L,i}}{2}- \frac{N_y\Tilde{\varpi}_{1,i}}{2})/N_y)} \cdot
\frac{\sin(\pi q_i-\frac{N_zv_{L,i}}{2}-\frac{N_z\Tilde{\varpi}_{2,i}}{2})}{\sin((\pi q_i-\frac{N_zv_{L,i}}{2}-\frac{N_z\Tilde{\varpi}_{2,i}}{2})/N_z)},
  \end{align}
\begin{comment}
After conducting the angle rotation operation, $\bar{y}_{n_zn_y}(t)$ in \eqref{1e11} and $\bar{r}(t)$ in \eqref{1e13} could be rewritten as
\begin{align}\label{1e15}
\breve{y}_{n_zn_y}(t) &= \bar{y}_{n_zn_y}(t)e^{-j\Tilde{\varpi}_1(n_y-1)}
 = \sum^{M_y}_{n_y=1}\hat{y}_{n_zn_y}(t)e^{j2\pi m_y(n_y-1)/M_y}e^{-j\Tilde{\varpi}_1(n_y-1)}\nonumber\\
  &=e^{-j2\pi v_{L,i}}\sqrt{p}s(t)\sum^{M_y}_{n_y=1}e^{-j2\pi(\frac{ d_1\cos\theta_{L,i}\cos\phi_{L,i}}{\lambda_c}-\frac{m_y}{M_y}+\frac{\Tilde{\varpi}_1}{2\pi})(n_y-1)},
  \end{align}
  \begin{align}\label{1e16}
\breve{r}(t)  &=\bar{r}(t)e^{-j\Tilde{\varpi}_{2,i}(n_z-1)}
= \sum^{M_z}_{n_z=1}r(t)e^{j2\pi m_z(n_z-1)/M_z}e^{-j\Tilde{\varpi}_{2,i}(n_z-1)}\nonumber\\
&=M_y\sqrt{p}s(t)\sum^{M_z}_{n_z=1} e^{-j2\pi (\frac{ d_1\sin\phi_{L,i}}{\lambda_c}-\frac{m_z}{M_z}+\frac{\Tilde{\varpi}_{2,i}}{2\pi})(n_z-1)},
\end{align}
where
\end{comment}
where $\Tilde{\varpi}_{1,i}$ and $\Tilde{\varpi}_{2,i}$ could be optimized with the one-dimensional search. Then,  we could obtain the estimated results as
  \begin{align}\label{1e18}
  \cos\hat{\Theta}_{L,i}\cos\hat{\Phi}_{L,i}=\frac{\lambda_c b_{n_i}}{N_y d_r }-\frac{\lambda_c\Tilde{\varpi}_{1,i}}{2\pi d_r},\nonumber\\
\sin\hat{\Phi}_{L,i}=\frac{\lambda_c q_{n_i}}{N_z d_r }-\frac{\lambda_c\Tilde{\varpi}_{2,i}}{2\pi d_r},
\end{align}
where $\hat{\Theta}_{L,i}$ and $\hat{\Phi}_{L,i}$ denote the final estimated azimuth and elevation AoAs after angle rotation, respectively. Furthermore, $\hat{\Theta}_{L,i}$ and $\hat{\Phi}_{L,i}$ could be expressed as
  \begin{align}\label{1em18}
\hat{\Phi}_{L,i}&=\arcsin\bigg(\frac{\lambda_c b_{n_i}}{N_z d_r }-\frac{\lambda_c\Tilde{\varpi}_{2,i}}{2\pi d_r}\bigg),\nonumber\\
\hat{\Theta}_{L,i} &= \arccos\bigg(\bigg(\frac{\lambda_c b_{n_i}}{N_y d_r }-\frac{\lambda_c\Tilde{\varpi}_{1,i}}{2\pi d_r}\bigg)\bigg/\sqrt{\bigg(1-\bigg(\frac{\lambda_c q_{n_i}}{N_z d_r }-\frac{\lambda_c\Tilde{\varpi}_{2,i}}{2\pi d_r}\bigg)^2\bigg)}\bigg).
\end{align}
\section{Derivation of the PDF of Angle Estimation Error} \label{PDF}
In this section, as the second step of the proposed framework, we derive the PDF of the angle estimation error, which will be used for deriving the variance of the angle estimation error.  To obtain the PDF of the angle estimation error, the first step is to derive the PDF of $\tilde{\Phi}_{L,i}$ based on the estimated angle $\hat{\Phi}_{L,i}$ and the property of the 2D-DFT. Then, the second step is to design an algorithm by deriving the PDF of $\tilde{\Theta}_{L,i}$ based on the estimated angle $\hat{\Theta}_{L,i}$ and the property of the 2D-DFT. The details are given as follows:
\subsection{PDF of $\tilde{\Phi}_{L,i}$}
Based on the above section, we can find that $\Tilde{\varpi}_{2,i}$ could be obtained by the one-dimensional search in the interval of
 $[-\frac{\pi}{N_z},\frac{\pi}{N_z}]$. We assume that there are $S_{2,i}$ grids points in the interval $[-\frac{\pi}{N_z},\frac{\pi}{N_z}]$
  and $s_{2,i} \in \{1,\cdots,S_{2,i}\}$ is the optimal point. Therefore, the optimal solution for the one-dimensional search is
 $\Tilde{\varpi}_{2,i}=\frac{2\pi s_{2,i}}{N_zS_{2,i}}$, and the estimation of $\sin{\Phi}_{L,i}$ is thus written as
\begin{align}\label{e9}
\sin\hat{\Phi}_{L,i} = \frac{\lambda_c q_{n_i}}{N_z d_r }-\frac{\lambda_c\Tilde{\varpi}_{2,i}}{2\pi d_r}
=\frac{\lambda_c q_{n_i}}{N_z d_r }-\frac{\lambda_c s_{2,i}}{N_z d_rS_{2,i}}.
\end{align}

For notation simplicity, let us define $\hat {Y_i}=\sin\hat{\Phi}_{L,i}$ and $Y_i=\sin\Phi_{L,i}$. Then, the estimation of $\Phi_{L,i}$ could be expressed as
\begin{align}\label{e10}
\hat{\Phi}_{L,i} = \arcsin \hat{Y_i}=\arcsin\left( \frac{\lambda_c q_{n_i}}{N_z d_r }-\frac{\lambda_c s_{2,i}}{N_z d_rS_{2,i}} \right).
\end{align}
According to \eqref{e9} and the property of the one-dimensional search
method, the value of $Y_i$ follows the uniform distribution within the region of $\left[\hat{Y_i}-a_i,\hat{Y_i}+a_i\right]$, where $a_i=\frac{\lambda_c}{2N_z d_rS_{2,i}}$, which is given by
\begin{align}\label{e11}
f_{Y_i}(y)=\left\{
\begin{array}{rcl}
\frac{1}{2a_i},& & {\hat{Y_i}-a_i\leq y \leq \hat{Y_i}+a_i}\\
0, & & {\rm others}.
\end{array} \right.
\end{align}
By denoting the estimation error of $Y_i$ as $\tilde{Y_i}$, we have $\hat{Y_i}=Y_i+\tilde{Y_i}$. Then, the distribution of $\tilde{Y_i}$ is given by
\begin{align}\label{e12}
f_{\tilde{Y_i}}(\tilde{y})=\left\{
\begin{array}{rcl}
\frac{1}{2a_i},& & {-a_i\leq \tilde{y} \leq a_i}\\
0, & & {\rm others.}
\end{array} \right.
\end{align}
Since $\Phi_{L,i}=\arcsin Y_i$, the cumulative density function (CDF) of $\Phi_{L,i}$ is derived as
\begin{align}\label{e13}
F_{\Phi_{L,i}}(\phi_{L,i})=\textrm{Pr}(\Phi_{L,i}\leq\phi_{L,i})=\textrm{Pr}(\arcsin Y_i \leq \phi_{L,i})=\textrm{Pr}(Y_i\leq\sin \phi_{L,i}).
\end{align}
By assuming that $\Phi_{L,i}\in(-\frac{\pi}{2},\frac{\pi}{2})$,
 the PDF of $\Phi_{L,i}$ could be derived as
\begin{align}\label{e14}
f_{\Phi_{L,i}}(\phi_{L,i}) &= \frac{\partial F_{\Phi_{L,i}}(\phi_{L,i})}{\partial\phi_{L,i}}
=\cos\phi_{L,i} f_{Y_i}(\sin \phi_{L,i})\nonumber\\
&=\left\{
\begin{array}{rcl}
\frac{\cos\phi_{L,i}}{2a_i},& & {\arcsin(\hat{Y_i}-a_i) \leq\phi_{L,i}\leq \arcsin(\hat{Y_i}+a_i)}\\
0, & & { \rm others.}
\end{array} \right.
\end{align}

As we have $\hat{\Phi}_{L,i}=\Phi_{L,i}+\tilde{\Phi}_{L,i}$, the CDF of the estimation error $\tilde{\Phi}_{L,i}$ is calculated as
\begin{align}\label{e15}
F_{\tilde{\Phi}_{L,i}}\left(\tilde{\phi}_{L,i}\right)&=\textrm{Pr}\left(\tilde{\Phi}_{L,i}\leq\tilde{\phi}_{L,i}\right)
=\textrm{Pr}\left(\hat{\Phi}_{L,i}-\Phi_{L,i}\leq\tilde{\phi}_{L,i}\right)\nonumber\\
&=\textrm{Pr}\left(\Phi_{L,i}\geq\hat{\Phi}_{L,i}-\tilde{\phi}_{L,i}\right)
=1-\textrm{Pr}\left(\Phi_{L,i}\leq\hat{\Phi}_{L,i}-\tilde{\phi}_{L,i}\right).
\end{align}
Define $a_{1,i}=\arcsin\left(\hat{Y_i}+a_i\right)$ and $a_{2,i}=\arcsin\left(\hat{Y_i}-a_i\right)$. Based on \eqref{e15}, the PDF of $\tilde{\Phi}_{L,i}$ is written as
\begin{align}\label{e16}
f_{\tilde{\Phi}_{L,i}}\left(\tilde{\phi}_{L,i}\right)&=\frac{\partial F_{\tilde{\Phi}_{L,i}}\left(\tilde{\phi}_{L,i}\right)}{\partial\tilde{\phi}_{L,i}}
=f_{{\Phi}_{L,i}}\left(\hat{\Phi}_{L,i}-\tilde{\phi}_{L,i}\right)\nonumber\\
&=\left\{
\begin{array}{rcl}
\frac{\cos\left(\hat{\Phi}_{L,i}-\tilde{\phi}_{L,i}\right)}{2a_i},& & {\hat{\Phi}_{L,i}-a_{1,i}\leq\tilde{\phi}_{L,i}\leq\hat{\Phi}_{L,i}-a_{2,i}}\\
0, & & {\rm others.}
\end{array} \right.
\end{align}

Moreover, as the estimation techniques are relatively mature, it is assumed that the estimation error is very small.
 Therefore, we have $\sin\left(\tilde{\phi}_{L,i}\right)\approx \tilde{\phi}_{L,i}$ and $\cos\left(\tilde{\phi}_{L,i}\right)\approx 1$.
  Consequently, we have the following approximation:
\begin{align}\label{1e16}
  \cos\left(\hat{\Phi}_{L,i}-\tilde{\phi}_{L,i}\right)=\cos\hat{\Phi}_{L,i}\cos\tilde{\phi}_{L,i}+\sin\hat{\Phi}_{L,i}
  \sin\tilde{\phi}_{L,i}
  \approx\cos\hat{\Phi}_{L,i}+\tilde{\phi}_{L,i}\sin\hat{\Phi}_{L,i}.
\end{align}
Then, the PDF of $\tilde{\Phi}_{L,i}$ is approximated as
\begin{align}\label{1e17}
f_{\tilde{\Phi}_{L,i}}\left(\tilde{\phi}_{L,i}\right)
\approx \left\{
\begin{array}{rcl}
\frac{\cos\hat{\Phi}_{L,i}+\tilde{\phi}_{L,i}\sin\hat{\Phi}_{L,i}}{2a_i},& & {\hat{\Phi}_{L,i}-a_{1,i}\leq\tilde{\phi}_{L,i}\leq\hat{\Phi}_{L,i}-a_{2,i}}\\
0, & & {\rm others.}
\end{array} \right.
\end{align}

\subsection{PDF of $\tilde{\Theta}_{L,i}$}
In this subsection, we derive the PDF of the estimation error $\tilde{\Theta}_{L,i}$.
 As the derivations are complicated, we summarize the main procedure in Algorithm \ref{alg:1}.

\begin{algorithm}
	\caption{Algorithm of deriving the PDF of $\tilde{\Phi}_{L,i}$}
	\label{alg:1}
	\begin{algorithmic}[1]
        \STATE Derive the PDF of $\cos\Phi_{L,i}$ by using the PDF of $\Tilde{Y_i}$ in \eqref{e12};
		\STATE Derive the PDF of $\cos\Theta_{L,i}$ for two cases by using the PDF of $\cos\Phi_{L,i}$ in \eqref{e24}
               and the PDF of $\cos\Phi_{L,i}\cos\Theta_{L,i}$ in \eqref{e25};
		\STATE Derive the PDF of $\tilde{\Theta}_{L,i}$ based on the two cases of the PDF of $\cos{\Theta}_{L,i}$
 in \eqref{e34} and \eqref{ee34}.
	\end{algorithmic}
\end{algorithm}

\subsubsection{PDF of $\cos\Phi_{L,i}$}
First of all, we need to derive the PDF of $\cos\Phi_{L,i}$, so that the PDF of $\cos\Theta_{L,i}$ could be calculated.

  Let ${X_i}=\cos\Phi_{L,i}$, which can be expressed as a function of $\Tilde{Y_i}$ as follows:
\begin{align}\label{e18}
{X_i} =\cos\Phi_{L,i}=\sqrt{1-\sin^2\Phi_{L,i}}= \sqrt{1-Y_i^2}=\sqrt{1-(\hat{Y_i}-\Tilde{Y_i})^2}.
\end{align}
By defining the estimation of ${X_i}$ as $\hat{{X_i}}=\cos\hat{\Phi}_{L,i}={X_i}+\Tilde{{X_i}}$, the estimation error $\Tilde{{X_i}}$ could be calculated as:
 \begin{align}\label{ee18}
\Tilde{{X_i}}=\hat{{X_i}}-\sqrt{1-(\hat{Y_i}-\Tilde{Y_i})^2}.
\end{align}
However, the expression \eqref{ee18} is complicated and thus challenging to derive a compact form of the PDF of $\Tilde{{X_i}}$.
Fortunately, since the value of $\Tilde{Y_i}$ is relatively small,
we can approximate $\Tilde{{X_i}}$ in \eqref{ee18} by using the Taylor expansion, which is given by
\begin{align}\label{e19}
\Tilde{{X_i}}=\sqrt{1-\hat{Y_i}^2}-\sqrt{1-(\hat{Y_i}-\Tilde{Y_i})^2}
 \approx \sqrt{1-\hat{Y_i}^2}-\bigg((1-\hat{Y_i}^2)^{\frac{1}{2}}+\frac{\Tilde{Y_i}\cdot\hat{Y_i}}{(1-\hat{Y_i}^2)^{\frac{1}{2}}}\bigg)
  = -\frac{\hat{Y_i}}{\hat{{X_i}}}\Tilde{Y_i}.
\end{align}
Utilizing \eqref{e19}, the CDF of $\Tilde{{X_i}}$ could be calculated as
\begin{align}\label{e21}
F_{\tilde{{X_i}}}(\tilde{x_i})=\textrm{Pr}(\tilde{{X_i}}\leq \Tilde{{x_i}})
\approx \textrm{Pr}\bigg(-\frac{\hat{Y_i}}{\hat{{X_i}}}\Tilde{Y_i} \leq \Tilde{{x_i}}\bigg)
= \textrm{Pr}\bigg(\Tilde{Y_i}\geq-\frac{\hat{{X_i}}}{\hat{Y_i}}\Tilde{{x_i}}\bigg)
=1-\textrm{Pr}\bigg(\Tilde{Y_i}\leq-\frac{\hat{{X_i}}}{\hat{Y_i}}\Tilde{{x_i}}\bigg).
\end{align}
Then, by using \eqref{e21}, the PDF of $\Tilde{{X_i}}$ could be calculated as
\begin{align}\label{e22}
f_{\tilde{{X_i}}}(\Tilde{{x_i}})=-\bigg(-\frac{\hat{{X_i}}}{\hat{Y_i}}\bigg)f_{\Tilde{Y_i}}\bigg(-\frac{\hat{{X_i}}}{\hat{Y_i}}\Tilde{{x_i}}\bigg)
=\left\{
\begin{array}{rcl}
\frac{\hat{{X_i}}}{2a_i\hat{Y_i}},& & {-\frac{\hat{Y_i}}{\hat{{X_i}}}a_i\leq\Tilde{{x_i}} \leq\frac{\hat{Y_i}}{\hat{{X_i}}}a_i}\\
0, & & {\rm others.}
\end{array} \right.
\end{align}
Furthermore, by using $\hat{{X_i}}={X_i}+\Tilde{{X_i}}$, the CDF of ${X_i}$ can be calculated as
\begin{align}\label{e23}
F_{X_i}({x_i})=\textrm{Pr}({X_i}\leq {x_i})= \textrm{Pr}(\hat{{X_i}}-\Tilde{{X_i}}\leq {x_i})= \textrm{Pr}(\Tilde{{X_i}}\geq \hat{{X_i}}-{x_i})=1-\textrm{Pr}(\tilde{{X_i}} \leq \hat{{X_i}}-{x_i}).
\end{align}

By using \eqref{e22} and \eqref{e23}, the PDF of ${X_i}$ could be written as
\begin{align}\label{e24}
f_{X_i}({x_i}) = \frac{\partial F_{X_i}({x_i})}{\partial {x_i}}=f_{\Tilde{{X_i}}}(\hat{{X_i}}-{x_i})=\left\{
\begin{array}{rcl}
\frac{\hat{{X_i}}}{2a_i\hat{Y_i}},& & {\hat{{X_i}}-\frac{\hat{Y_i}}{\hat{{X_i}}}a_i\leq {x_i} \leq\hat{{X_i}}-\frac{\hat{Y_i}}{\hat{{X_i}}}a_i}\\
0, & & {\rm others.}
\end{array} \right.
\end{align}
\subsubsection{PDF of $\cos\Theta_{L,i}$}
By using the PDF of $\cos\Phi_{L,i}$ and $\cos\Phi_{L,i}\cos\Theta_{L,i}$, we can derive the PDF of $\cos\Theta_{L,i}$ as follows.

Firstly, from the above section, we know that $\Tilde{\varpi}_{1,i}$ could be obtained by using the one-dimensional search in the interval of
 $[-\frac{\pi}{N_y},\frac{\pi}{N_y}]$. Similarly, we assume that there are $S_1$ grids points in the interval
  $[-\frac{\pi}{N_y},\frac{\pi}{N_y}]$
  and $s_{1,i}\in \{1,\cdots,S_{1,i}\}$ is the optimal point. Therefore, the optimal solution for the one-dimensional search is
   $\Tilde{\varpi}_{1,i}=\frac{2\pi s_{1,i}}{N_yS_{1,i}}$, and the estimated $\cos\Phi_{L,i}\cos\Theta_{L,i}$ is given by
\begin{align}\label{ew25}
\hat{Z} =  \cos\hat{\Theta}_{L,i}\cos\hat{\Phi}_{L,i}&=\frac{\lambda_c b_{n_i}}{N_y d_1 }-\frac{\lambda_c\Tilde{\varpi}_{1,i}}{2\pi d_r}
=\frac{\lambda_c b_{n_i}}{N_y d_r }-\frac{\lambda_c s_{1,i}}{N_y d_rS_{1,i}}.
\end{align}
By using \eqref{ew25} and the nature of the one-dimensional search
 method, the real value of ${Z_i}=\cos{\Theta}_{L,i}\cos{\Phi}_{L,i}$ follows the uniform distribution within the region of
 $\left[\hat{{Z_i}}-{b_i},\hat{{Z_i}}+{b_i}\right]$, where ${b_i}=\frac{\lambda_c}{2N_y d_1S_{1,i}}$. The PDF of ${Z_i}$ is thus given by
\begin{align}\label{e25}
f_{{Z_i}}( z_i)=\left\{
\begin{array}{rcl}
\frac{1}{2{b_i}},& & {\hat{{Z_i}}-{b_i}\leq z_i\leq\hat{{Z_i}}+{b_i}}\\
0, & & {\rm others.}
\end{array} \right.
\end{align}
By denoting the estimation error of ${Z_i}$ as $\Tilde{{Z_i}}$, we have $\hat{{Z_i}}={Z_i}+\Tilde{{Z_i}}$. Then, the distribution of $\Tilde{{Z_i}}$ is given by
\begin{align}\label{e26}
f_{\tilde{{Z_i}}}(\Tilde{z_i})=\left\{
\begin{array}{rcl}
\frac{1}{2{b_i}},& & {-{b_i}\leq\Tilde{z_i}\leq {b_i}}\\
0, & & {\rm others.}
\end{array} \right.
\end{align}

Next, for simplicity, let us denote ${U_i}=\cos\Theta_{L,i}$. Then, by utilizing the definition of ${Z_i}$ below \eqref{ew25}
and ${X_i}$ above \eqref{e18},  we have ${Z_i}={U_i}{X_i}$. By combining the PDF of ${X_i}$ in \eqref{e24} and the PDF of ${Z_i}$ in \eqref{e25},
 we could derive the PDF of ${U_i}$ as follows
\begin{align}\label{e27}
f_{{U_i}}(u_i)=\int_{\hat{{X_i}}-\frac{\hat{Y_i}}{\hat{{X_i}}}a_i}^{\hat{{X_i}}+\frac{\hat{Y_i}}{\hat{{X_i}}}a_i}
{x}f_{X_i}(x)f_{Z_i}(u_ix)dx.
\end{align}

According to \eqref{e25}, it is observed that $f_{Z_i}(u_ix)$ is non-zero when
$\hat{{Z_i}}-{b_i}\leq u_ix \leq \hat{{Z_i}}+{b_i}$, which determines the PDF of ${U_i}$. Therefore, we need to discuss the different conditions
according to the non-zero intervals of $f_{Z_i}(u_ix)$ and $f_{X_i}(x)$ in the following. For notation brevity, we denote
  ${\alpha_{1,i}} = \hat{{X_i}}-\frac{\hat{Y_i}}{\hat{{X_i}}}a_i$, ${\alpha_{2,i}}= \hat{{X_i}}+\frac{\hat{Y_i}}{\hat{{X_i}}}a_i$,
   ${\beta_{1,i}}=\hat{{Z_i}}-{b_i}$ and ${\beta_{2,i}}=\hat{{Z_i}}+{b_i}$.
\begin{comment}
{Case 1}: When $\hat{{X_i}}-\frac{\hat{Y}}{\hat{{X_i}}}a<\hat{{X_i}}+\frac{\hat{Y}}{\hat{{X_i}}}a<\frac{\hat{{Z_i}}-b}{u}<\frac{\hat{{Z_i}}+b}{u}$, we could derive the interval of $u$ as $(-\infty,\frac{\hat{{Z_i}}-b}{\hat{{X_i}}+\frac{\hat{Y}}{\hat{{X_i}}}a})$, and the PDF of $u$ in this interval is written as
\begin{align}\label{e28}
f_{ {U_i}}( u)=0.
\end{align}
\end{comment}

{Condition 1}: If ${\alpha_{1,i}}<\frac{{\beta_{1,i}}}{u_i}\leq{\alpha_{2,i}}<\frac{{\beta_{2,i}}}{u_i}$, the integral interval of $x$ in \eqref{e27}
can be recast as $\left[\frac{{\beta_{1,i}}}{u_i},{\alpha_{2,i}}\right]$, thereby yielding the PDF of ${U_i}$ as
\begin{align}\label{e29}
f_{{U_i}}(u_i)=\int_{\frac{{\beta_{1,i}}}{u_i}}^{{\alpha_{2,i}}}\frac{1}{2b_i}\cdot\frac{\hat{{X_i}}}{2a_i\hat{Y_i}}\cdot xdx
=\frac{\hat{{X_i}}}{4a_i{b_i}\hat{Y_i}}\int_{\frac{{\beta_{1,i}}}{u_i}}^{{\alpha_{2,i}}}xdx
=\frac{\hat{{X_i}}}{8a_i{b_i}\hat{Y_i}}\left({\alpha_{2,i}}^2-\left(\frac{{\beta_{1,i}}}{u_i}\right)^2\right).
\end{align}
Furthermore, the interval of $u_i$ is given by
\begin{align}\label{eq29}
\left[\frac{{\beta_{1,i}}}{{\alpha_{2,i}}}, \frac{{\beta_{1,i}}}{{\alpha_{1,i}}}\right)\cap\left(-\infty,\frac{{\beta_{2,i}}}{{\alpha_{2,i}}}\right).
\end{align}
Based on \eqref{eq29}, it is necessary to compare $\frac{{\beta_{1,i}}}{{\alpha_{1,i}}}$ with $\frac{{\beta_{2,i}}}{{\alpha_{2,i}}}$, so that the interval of $u_i$ can be further determined. If $\frac{{\beta_{2,i}}}{{\alpha_{2,i}}}>\frac{{\beta_{1,i}}}{{\alpha_{1,i}}}$ holds, the interval can be rewritten as $\left[\frac{{\beta_{1,i}}}{{\alpha_{2,i}}}, \frac{{\beta_{1,i}}}{{\alpha_{1,i}}}\right)$. Otherwise, the interval is $\left[\frac{{\beta_{1,i}}}{{\alpha_{2,i}}}, \frac{{\beta_{2,i}}}{{\alpha_{2,i}}}\right)$.

{Condition 2}: If $\frac{{\beta_{1,i}}}{u_i}\leq{\alpha_{1,i}}<\frac{{\beta_{2,i}}}{u_i}\leq{\alpha_{2,i}}$, the integral interval of $x$ in \eqref{e27}
can be recast as $\left[{\alpha_{1,i}},\frac{{\beta_{2,i}}}{u_i}\right]$, thus the PDF of ${U_i}$  is given by
\begin{align}\label{e30}
f_{{U_i}}(u_i)=\int_{{\alpha_{1,i}}}^{\frac{{\beta_{2,i}}}{u_i}}\frac{1}{2b_i}\cdot\frac{\hat{{X_i}}}{2a_i\hat{Y_i}}\cdot {x}d{x}
=\frac{\hat{{X_i}}}{4a_i{b_i}\hat{Y_i}}\int_{{\alpha_{1,i}}}^{\frac{{\beta_{2,i}}}{u_i}}{x}d{x}
=\frac{\hat{{X_i}}}{8a_i{b_i}\hat{Y_i}}\left(\left(\frac{{\beta_{2,i}}}{u_i}\right)^2-{\alpha_{1,i}}^2\right).
\end{align}
The interval of $u_i$ is given by
\begin{align}\label{eq30}
\left[\frac{{\beta_{2,i}}}{{\alpha_{2,i}}}, \frac{{\beta_{2,i}}}{{\alpha_{1,i}}}\right)\cap\left[\frac{{\beta_{1,i}}}{{\alpha_{1,i}}},+\infty\right).
\end{align}
As a result, if $\frac{{\beta_{2,i}}}{{\alpha_{2,i}}}>\frac{{\beta_{1,i}}}{{\alpha_{1,i}}}$ holds, the interval can be further recast as $\left[\frac{{\beta_{2,i}}}{{\alpha_{2,i}}}, \frac{{\beta_{2,i}}}{{\alpha_{1,i}}}\right)$. Otherwise, the interval can be derived as $\left[\frac{{\beta_{1,i}}}{{\alpha_{1,i}}}, \frac{{\beta_{2,i}}}{{\alpha_{1,i}}}\right)$.

{Condition 3}: If $\frac{{\beta_{1,i}}}{u_i}\leq{\alpha_{1,i}}<{\alpha_{2,i}}<\frac{{\beta_{2,i}}}{u_i}$, the integral interval of $x$ in \eqref{e27}
can be derived as $\left[{\alpha_{1,i}},{\alpha_{2,i}}\right]$. Thus, the PDF of ${U_i}$ is
\begin{align}\label{e31}\hspace{-2cm}
f_{{U_i}}(u_i)&=\int_{{\alpha_{1,i}}}^{{\alpha_{2,i}}}\frac{1}{2{b_i}}\cdot\frac{\hat{{X_i}}}{2a_i\hat{Y_i}}\cdot xdx
=\frac{\hat{{X_i}}}{4a_i{b_i}\hat{Y_i}}\int_{{\alpha_{1,i}}}^{{\alpha_{2,i}}}xdx
=\frac{\hat{{X_i}}}{8a_i{b_i}\hat{Y_i}}\left({\alpha_{2,i}}^2-{\alpha_{1,i}}^2\right)\nonumber\\
&=\frac{\hat{{X_i}}}{8a_i{b_i}\hat{Y_i}}\left(\left(\hat{{X_i}}+\frac{\hat{Y_i}}{\hat{{X_i}}}a_i\right)^2-\left(\hat{{X_i}}-\frac{\hat{Y_i}}{\hat{{X_i}}}a_i\right)^2\right)
=\frac{\hat{{X_i}}}{2{b_i}}.
\end{align}
The interval of $u_i$ is accordingly given by $\left[\frac{{\beta_{1,i}}}{{\alpha_{1,i}}},\frac{{\beta_{2,i}}}{{\alpha_{2,i}}}\right)$.

{Condition 4}: If ${\alpha_{1,i}}<\frac{{\beta_{1,i}}}{u_i}<\frac{{\beta_{2,i}}}{u_i}\leq{\alpha_{2,i}}$, the integral interval of $x$ in \eqref{e27}
can be derived as $\left[\frac{{\beta_{1,i}}}{u_i},\frac{{\beta_{2,i}}}{{U_i}}\right]$. Thus, the PDF of ${U_i}$ is
\begin{align}\label{e32}
f_{{U_i}}(u_i)=\int_{\frac{{\beta_{1,i}}}{u_i}}^{\frac{{\beta_{2,i}}}{u_i}}\frac{\hat{{X_i}}}{4a_i{b_i}\hat{Y_i}}\cdot xdx
=\frac{\hat{{X_i}}}{4a_i{b_i}\hat{Y_i}}\int_{\frac{{\beta_{1,i}}}{u_i}}^{\frac{{\beta_{2,i}}}{u_i}}xdx
=\frac{\hat{{X_i}}}{8a_i{b_i}\hat{Y_i}}\left(\left(\frac{{\beta_{2,i}}}{u_i}\right)^2-\left(\frac{{\beta_{1,i}}}{u_i}\right)^2\right)
=\frac{\hat{{X_i}}\hat{{Z_i}}}{2a_i\hat{Y_i}u_i^2}.
\end{align}
Additionally, after some mathematical manipulations, we can derive the interval of $u_i$ as $\left[\frac{{\beta_{2,i}}}{{\alpha_{2,i}}},\frac{{\beta_{1,i}}}{{\alpha_{1,i}}}\right)$.

\begin{comment}
{Case 6}: When $\frac{\hat{{Z_i}}-b}{u}<\frac{\hat{{Z_i}}+b}{u}\leq\hat{{X_i}}-\frac{\hat{Y}}{\hat{{X_i}}}a<\hat{{X_i}}+\frac{\hat{Y}}{\hat{{X_i}}}a$, we have the interval of $u$ as  $[\frac{\hat{{Z_i}}+b}{\hat{{X_i}}-\frac{\hat{Y}}{\hat{{X_i}}}a},+\infty)$, the PDF of $u$ is written as
\begin{align}\label{e33}
f_{{U_i}}(u)=0.
\end{align}
\end{comment}
Based on the above discussions, by comparing $\frac{{\beta_{2,i}}}{{\alpha_{2,i}}}$ with $\frac{{\beta_{1,i}}}{{\alpha_{1,i}}}$, the PDF of ${U_i}$ can be simplified as the following two cases:

Case 1: If $\frac{{\beta_{2,i}}}{{\alpha_{2,i}}}>\frac{{\beta_{1,i}}}{{\alpha_{1,i}}}$ holds, the intervals of $u$ in
  Condition 1 and Condition 2 are given by $\left[\frac{{\beta_{1,i}}}{{\alpha_{2,i}}}, \frac{{\beta_{1,i}}}{{\alpha_{1,i}}}\right)$
and $\left[\frac{{\beta_{2,i}}}{{\alpha_{2,i}}},\frac{{\beta_{2,i}}}{{\alpha_{1,i}}}\right)$.
Additionally, Condition 3 is valid, whereas Condition 4 is invalid. Therefore, the PDF of ${U_i}$ is written as
\begin{align}\label{e34}
f_{{U_i}}(u_i)=\left\{
\begin{array}{rcl}
\frac{\hat{{X_i}}}{8a_ib_i\hat{Y_i}}\left({\alpha_{2,i}}^2-\left(\frac{{\beta_{1,i}}}{u_i}\right)^2\right),
& & {\frac{{\beta_{1,i}}}{{\alpha_{2,i}}}\leq u_i<\frac{{\beta_{1,i}}}{{\alpha_{1,i}}}}\\
\frac{\hat{{X_i}}}{2b_i},
& & {\frac{{\beta_{1,i}}}{{\alpha_{1,i}}}\leq u_i<\frac{{\beta_{2,i}}}{{\alpha_{2,i}}}}\\
\frac{\hat{{X_i}}}{8a_ib_i\hat{Y_i}}\left(\left(\frac{{\beta_{2,i}}}{u_i}\right)^2-{\alpha_{1,i}}^2\right), & &
{\frac{{\beta_{2,i}}}{{\alpha_{2,i}}}\leq u_i\leq \frac{{\beta_{2,i}}}{{\alpha_{1,i}}}}\\
0, & & {\rm others.}
\end{array} \right.
\end{align}

Case 2: If $\frac{{\beta_{2,i}}}{{\alpha_{2,i}}}<\frac{{\beta_{1,i}}}{{\alpha_{1,i}}}$ holds,
 the intervals of $u_i$ in  Condition 1 and Condition 2
are given by $\left[\frac{{\beta_{1,i}}}{{\alpha_{2,i}}}, \frac{{\beta_{2,i}}}{{\alpha_{2,i}}}\right)$
and $\left[\frac{{\beta_{1,i}}}{{\alpha_{1,i}}},\frac{{\beta_{2,i}}}{{\alpha_{1,i}}}\right)$, respectively.
Moreover, Condition 4 is valid, while Condition 3 is invalid.
Accordingly, the PDF of ${U_i}$ can be derived as
\begin{align}\label{ee34}
f_{{U_i}}(u_i)=\left\{
\begin{array}{rcl}
\frac{\hat{{X_i}}}{8a_ib_i\hat{Y_i}}\left({\alpha_{2,i}}^2-\left(\frac{{\beta_{1,i}}}{u_i}\right)^2\right),& &
 {\frac{{\beta_{1,i}}}{{\alpha_{2,i}}}\leq u_i<\frac{{\beta_{2,i}}}{{\alpha_{2,i}}}}\\
\frac{\hat{{X_i}}\hat{{Z_i}}}{2a_i\hat{Y_i}u_i^2}, & &
 {\frac{{\beta_{2,i}}}{{\alpha_{2,i}}}\leq u_i<\frac{{\beta_{1,i}}}{{\alpha_{1,i}}}}\\
\frac{\hat{{X_i}}}{8a_ib_i\hat{Y_i}}\left(\left(\frac{{\beta_{2,i}}}{u_i}\right)^2-{\alpha_{1,i}}^2\right),
 & & {\frac{{\beta_{1,i}}}{{\alpha_{1,i}}}\leq u_i\leq\frac{{\beta_{2,i}}}{{\alpha_{1,i}}}}\\
0, & & {\rm others.}
\end{array} \right.
\end{align}
\subsubsection{PDF of $\Theta_{L,i}$}
By using the PDF of $\cos\Theta_{L,i}$, the PDF of $\Theta_{L,i}$ can be derived as follows.

First of all, the CDF of $\Theta_{L,i}$ can be derived as follows
\begin{align}\label{e36}
F_{\Theta_{L,i}}(\theta_{L,i})=\textrm{Pr}(\Theta_{L,i}\leq\theta_{L,i})=\textrm{Pr}(\arccos {U_i} \leq \theta_{L,i})=1-\textrm{Pr}({U_i}\leq\cos\theta_{L,i}).
\end{align}
Then, the PDF of $\Theta_{L,i}$ is obtained as follows
\begin{align}\label{e37}
f_{\Theta_{L,i}}(\theta_{L,i})=\frac{\partial F_{\Theta_{L,i}}(\theta_{L,i})}{\partial \theta_{L,i}}
=\sin\theta_{L,i}f_{U_i}(\cos\theta_{L,i}).
\end{align}

Furthermore, we consider the indoor positioning system in this paper, where the RISs are supposed to be mounted on the wall.
Hence we have $\Theta_{L,i}\in(0,\pi)$.  Thus, $\Theta_{L,i}$ decreases monotonically with ${U_i}$. According to the PDF of ${U_i}$ in the aforementioned two cases, we can derive the PDF of $\Theta_{L,i} $ in the following.

Case 1: When $\frac{{\beta_{2,i}}}{{\alpha_{2,i}}}>\frac{{\beta_{1,i}}}{{\alpha_{1,i}}}$, according to \eqref{e34}, we can derive the PDF of $\Theta_{L,i} $, which is given by
\begin{align}\label{e38}
f_{\Theta_{L,i}}( \theta_{L,i})=\left\{
\begin{array}{rcl}
\frac{\hat{{X_i}}\sin\theta_{L,i}}{8a_ib_i\hat{Y_i}}\left[\left(\frac{{\beta_{2,i}}}{\cos\theta_{L,i}}\right)^2-{\alpha_{1,i}}^2\right],
 & & {\arccos\frac{{\beta_{2,i}}}{{\alpha_{1,i}}}\leq\theta_{L,i}\leq\arccos\frac{{\beta_{2,i}}}{{\alpha_{2,i}}}}\\
\frac{\hat{{X_i}}\sin\theta_{L,i}}{2b_i}, & &
 {\arccos\frac{{\beta_{2,i}}}{{\alpha_{2,i}}}<\theta_{L,i}\leq\arccos\frac{{\beta_{1,i}}}{{\alpha_{1,i}}}}\\
\frac{\hat{{X_i}}\sin\theta_{L,i}}{8a_ib_i\hat{Y_i}}\left[{\alpha_{2,i}}^2-(\frac{{\beta_{1,i}}}{\cos\theta_{L,i}})^2\right],
& & {\arccos\frac{{\beta_{1,i}}}{{\alpha_{1,i}}}<\theta_{L,i}\leq\arccos\frac{{\beta_{1,i}}}{{\alpha_{2,i}}}}\\
0, & & {\rm others.}
\end{array} \right.
\end{align}

Case 2: When $\frac{{\beta_{2,i}}}{{\alpha_{2,i}}}<\frac{{\beta_{1,i}}}{{\alpha_{1,i}}}$, we can derive the PDF of $\Theta_{L,i} $ based on \eqref{ee34}, which is given by
\begin{align}\label{e338}
f_{\Theta_{L,i}}( \theta_{L,i})=\left\{
\begin{array}{rcl}
\frac{\hat{{X_i}}\sin\theta_{L,i}}{8a_ib_i\hat{Y_i}}\left[\left(\frac{{\beta_{2,i}}}{\cos\theta_{L,i}}\right)^2-{\alpha_{1,i}}^2\right],
 & & {\arccos\frac{{\beta_{2,i}}}{{\alpha_{1,i}}}\leq\theta_{L,i}\leq\arccos\frac{{\beta_{1,i}}}{{\alpha_{1,i}}}}\\
\frac{\hat{{X_i}}\hat{{Z_i}}\sin\theta_{L,i}}{2a_i\hat{Y_i}(\cos\theta_{L,i})^2},
 & & {\arccos\frac{{\beta_{1,i}}}{{\alpha_{1,i}}}\leq\theta_{L,i}<\arccos\frac{{\beta_{2,i}}}{{\alpha_{2,i}}}}\\
\frac{\hat{{X_i}}\sin\theta_{L,i}}{8a_ib_i\hat{Y_i}}\left[{\alpha_{2,i}}^2-(\frac{{\beta_{1,i}}}{\cos\theta_{L,i}})^2\right],
& & {\arccos\frac{{\beta_{2,i}}}{{\alpha_{2,i}}}\leq\theta_{L,i}<\arccos\frac{{\beta_{1,i}}}{{\alpha_{2,i}}}}\\
0, & & {\rm others.}
\end{array} \right.
\end{align}
\subsubsection{PDF of $\tilde{\Theta}_{L,i}$}
By using the CDF and PDF of $\Theta_{L,i}$ in \eqref{e36}, \eqref{e38} and \eqref{e338}, we can calculate the CDF and PDF of $\tilde{\Theta}_{L,i}$ as follows.

Firstly, based on $\hat{\Theta}_{L,i}={\Theta}_{L,i}+\tilde{\Theta}_{L,i}$, the CDF of $\tilde{\Theta}_{L,i}$ can be derived as follows
\begin{align}\label{e39}
F_{\tilde{\Theta}_{L,i}}(\tilde{\theta}_{L,i})&=\textrm{Pr}(\tilde{\Theta}_{L,i}\leq\tilde{\theta}_{L,i})
=\textrm{Pr}(\hat{\Theta}_{L,i}-\Theta_{L,i}\leq\tilde{\theta}_{L,i})\nonumber\\
&=\textrm{Pr}(\Theta_{L,i}\geq\hat{\Theta}_{L,i}-\tilde{\theta}_{L,i})=
1-\textrm{Pr}(\Theta_{L,i}\leq\hat{\Theta}_{L,i}-\tilde{\theta}_{L,i}).
\end{align}

Then, based on the above two cases of $f_{\Theta_{L,i}}( \theta_{L,i})$ in \eqref{e38} and \eqref{e338}, the PDF of $\tilde{\Theta}_{L,i}$ can be derived accordingly by using \eqref{e39}.

Case 1: When $\frac{{\beta_{2,i}}}{{\alpha_{2,i}}}>\frac{{\beta_{1,i}}}{{\alpha_{1,i}}}$, by using \eqref{e38}, the PDF of $\tilde{\Theta}_{L,i}$ is calculated as
\begin{align}\label{eq40}
&f_{\tilde{\Theta}_{L,i}}(\tilde{\theta}_{L,i})=\frac{\partial F_{\tilde{\Theta}_{L,i}}(\tilde{\theta}_{L,i})}{\partial\tilde{\theta}_{L,i}}
=f_{\Theta_{L,i}}(\hat{\Theta}_{L,i}-\tilde{\theta}_{L,i})\nonumber\\
&=\left\{
\begin{array}{rcl}
\frac{\hat{{X_i}}\sin(\hat{\Theta}_{L,i}-\tilde{\theta}_{L,i})}{8a_ib_i\hat{Y_i}}
\left[{\alpha_{2,i}}^2-(\frac{{\beta_{1,i}}}{\cos(\hat{\Theta}_{L,i}-\tilde{\theta}_{L,i})})^2\right],
& & {\hat{\Theta}_{L,i}-\arccos\frac{{\beta_{1,i}}}{{\alpha_{2,i}}}
\leq\tilde{\theta}_{L,i}<\hat{\Theta}_{L,i}-\arccos\frac{{\beta_{1,i}}}{{\alpha_{1,i}}}}\\
\frac{\hat{{X_i}}\sin(\hat{\Theta}_{L,i}-\tilde{\theta}_{L,i})}{2b_i}, & &
 {\hat{\Theta}_{L,i}-\arccos\frac{{\beta_{1,i}}}{{\alpha_{1,i}}}\leq\tilde{\theta}_{L,i}<
 \hat{\Theta}_{L,i}-\arccos\frac{{\beta_{2,i}}}{{\alpha_{2,i}}}}\\
 \frac{\hat{{X_i}}\sin(\hat{\Theta}_{L,i}-\tilde{\theta}_{L,i})}{8a_ib_i\hat{Y_i}}
\left[\left(\frac{{\beta_{2,i}}}{\cos(\hat{\Theta}_{L,i}-\tilde{\theta}_{L,i})}\right)^2-{\alpha_{1,i}}^2\right],
 & & {\hat{\Theta}_{L,i}-\arccos\frac{{\beta_{2,i}}}{{\alpha_{2,i}}}
\leq\tilde{\theta}_{L,i}\leq\hat{\Theta}_{L,i}-\arccos\frac{{\beta_{2,i}}}{{\alpha_{1,i}}}}\\
0, & & {\rm others.}
\end{array} \right.
\end{align}
Case 2: When $\frac{{\beta_{2,i}}}{{\alpha_{2,i}}}<\frac{{\beta_{1,i}}}{{\alpha_{1,i}}}$, we can derive the PDF of $\tilde{\Theta}_{L,i}$ by using \eqref{e338}, which is given by
\begin{align}\label{eeq40}
&f_{\tilde{\Theta}_{L,i}}(\tilde{\theta}_{L,i})=\frac{\partial F_{\tilde{\theta}_{L,i}}(\tilde{\theta}_{L,i})}{\partial\tilde{\theta}_{L,i}}
=f_{\Theta_{L,i}}(\hat{\Theta}_{L,i}-\tilde{\theta}_{L,i})\nonumber\\
&=\left\{
\begin{array}{rcl}
\frac{\hat{{X_i}}\sin(\hat{\Theta}_{L,i}-\tilde{\theta}_{L,i})}{8a_ib_i\hat{Y_i}}
\left({\alpha_{2,i}}^2-\left(\frac{{\beta_{1,i}}}{\cos(\hat{\Theta}_{L,i}-\tilde{\theta}_{L,i})}\right)^2\right],
& & {\hat{\Theta}_{L,i}-\arccos\frac{{\beta_{1,i}}}{{\alpha_{2,i}}}
\leq\tilde{\theta}_{L,i}<\hat{\Theta}_{L,i}-\arccos\frac{{\beta_{2,i}}}{{\alpha_{2,i}}}}\\
\frac{\hat{{X_i}}\hat{{Z_i}}\sin(\hat{\Theta}_{L,i}-\tilde{\theta}_{L,i})}{2a_i\hat{Y_i}[\cos(\hat{\Theta}_{L,i}-\tilde{\theta}_{L,i})]^2}, & &
  {\hat{\Theta}_{L,i}-\arccos\frac{{\beta_{2,i}}}{{\alpha_{2,i}}}\leq\tilde{\theta}_{L,i}<
 \hat{\Theta}_{L,i}-\arccos\frac{{\beta_{1,i}}}{{\alpha_{1,i}}}}\\
 \frac{\hat{{X_i}}\sin(\hat{\Theta}_{L,i}-\tilde{\theta}_{L,i})}{8a_ib_i\hat{Y_i}}
\left[\left(\frac{{\beta_{2,i}}}{\cos(\hat{\Theta}_{L,i}-\tilde{\theta}_{L,i})}\right)^2-{\alpha_{1,i}}^2\right],
 & & {\hat{\Theta}_{L,i}-\arccos\frac{{\beta_{1,i}}}{{\alpha_{1,i}}}
\leq\tilde{\theta}_{L,i}\leq\hat{\Theta}_{L,i}-\arccos\frac{{\beta_{2,i}}}{{\alpha_{1,i}}}}\\
0, & & {\rm others.}
\end{array} \right.
\end{align}

To facilitate the error analysis, we now aim to derive the approximation of $f_{\tilde{\Theta}_{L,i}}(\tilde{\theta}_{L,i})$ in this paper.
 As we have assumed that the estimation error is very small,
  we have $\sin\tilde{\theta}_{L,i}\approx\tilde{\theta}_{L,i}$ and $\cos\tilde{\theta}_{L,i}\approx1$, leading to
\begin{align}\label{e41}
  \sin(\hat{\Theta}_{L,i}-\tilde{\theta}_{L,i})=\sin\hat{\Theta}_{L,i}\cos\tilde{\theta}_{L,i}-\sin\tilde{\theta}_{L,i}\cos\hat{\Theta}_{L,i} \approx\sin\hat{\Theta}_{L,i}-\tilde{\theta}_{L,i}\cos\hat{\Theta}_{L,i}, \nonumber \\
  \cos(\hat{\Theta}_{L,i}-\tilde{\theta}_{L,i})=\cos\hat{\Theta}_{L,i}\cos\tilde{\theta}_{L,i}+\sin\hat{\Theta}_{L,i}\sin\tilde{\theta}_{L,i}
  \approx\cos\hat{\Theta}_{L,i}+\tilde{\theta}_{L,i}\sin\hat{\Theta}_{L,i}.
\end{align}
Using the approximations in \eqref{e41}, we can derive the approximation of $f_{\tilde{\Theta}_{L,i}}(\tilde{\theta}_{L,i})$ according to the above two cases. Furthermore, by denoting $\sin\hat{\Theta}_{L,i} = \hat{V_i}$ , $\cos\hat{\Theta}_{L,i} = \hat{{U_i}}$,
 $B_{1,i}=\hat{\Theta}_{L,i}-\arccos\frac{{\beta_{1,i}}}{{\alpha_{2,i}}}$,
$B_{2,i}=\hat{\Theta}_{L,i}-\arccos\frac{{\beta_{1,i}}}{{\alpha_{1,i}}}$,
$B_{3,i}=\hat{\Theta}_{L,i}-\arccos\frac{{\beta_{2,i}}}{{\alpha_{2,i}}}$,
$B_{4,i}=\hat{\Theta}_{L,i}-\arccos\frac{{\beta_{2,i}}}{{\alpha_{1,i}}}$,
  the expression could be further simplified in the following.

\begin{comment}&\approx\left\{
\begin{array}{rcl}
\frac{\hat{{X_i}}(\sin\hat{\Theta}_{L,i}-\tilde{\theta}_{L,i}\cos\hat{\Theta}_{L,i})}{8ab\hat{Y_i}}
\left[{\alpha_{2,i}}^2-\left(\frac{{\beta_{1,i}}}{\cos\hat{\Theta}_{L,i}+\tilde{\theta}_{L,i}\sin\hat{\Theta}_{L,i}}\right)^2\right],&& {B_{1,i}\leq\tilde{\theta}_{L,i}< B_{2,i}}\\
\frac{\hat{{X_i}}(\sin\hat{\Theta}_{L,i}-\tilde{\theta}_{L,i}\cos\hat{\Theta}_{L,i})}{2b}, &&
 {B_{2,i}\leq\tilde{\theta}_{L,i}< B_{3,i}}\\
 \frac{\hat{{X_i}}(\sin\hat{\Theta}_{L,i}-\tilde{\theta}_{L,i}\cos\hat{\Theta}_{L,i})}{8ab\hat{Y_i}}
\left[\left(\frac{{\beta_{2,i}}}{\cos\hat{\theta}_{L,i}+\Delta\theta_{L,i}\sin\hat{\theta}_{L,i}}\right)^2-{\alpha_{1,i}}^2\right], && {B_{3,i}\leq\tilde{\theta}_{L,i}\leq B_{4,i}}\\
0, && {\rm others.}
\end{array} \right.\nonumber\\\end{comment}
{Case 1}: When $\frac{{\beta_{2,i}}}{{\alpha_{2,i}}}>\frac{{\beta_{1,i}}}{{\alpha_{1,i}}}$, based on \eqref{eq40}, we can derive the approximation of $f_{\tilde{\Theta}_{L,i}}(\tilde{\theta}_{L,i})$, which is written as
\begin{align} \label{e40}
f_{\tilde{\Theta}_{L,i}}(\tilde{\theta}_{L,i})
\approx\left\{
\begin{array}{rcl}
	\frac{\hat{{X_i}}(\hat{V_i}-\tilde{\theta}_{L,i}\hat{{U_i}})}{8a_ib_i\hat{Y_i}}
	\left[{\alpha_{2,i}}^2-\left(\frac{{\beta_{1,i}}}{\hat{{U_i}}+\tilde{\theta}_{L,i}\hat{V_i}}\right)^2\right],& & {B_{1,i}\leq\tilde{\theta}_{L,i}< B_{2,i}}\\
	\frac{\hat{{X_i}}(\hat{V_i}-\tilde{\theta}_{L,i}\hat{{U_i}})}{2b_i}, & &
	{B_{2,i}\leq\tilde{\theta}_{L,i}< B_{3,i}}\\
	\frac{\hat{{X_i}}(\hat{V_i}-\tilde{\theta}_{L,i}\hat{{U_i}})}{8a_ib_i\hat{Y_i}}
	\left[\left(\frac{{\beta_{2,i}}}{\hat{{U_i}}+\tilde{\theta}_{L,i}\hat{V_i}}\right)^2-{\alpha_{1,i}}^2\right], & & {B_{3,i}\leq\tilde{\theta}_{L,i}\leq B_{4,i}}\\
	0, & & {\rm others.}
\end{array} \right.
\end{align}

\begin{comment}&\approx\left\{
\begin{array}{rcl}
\frac{\hat{{X_i}}(\sin\hat{\Theta}_{L,i}-\tilde{\theta}_{L,i}\cos\hat{\Theta}_{L,i})}{8ab\hat{Y_i}}
\left[{\alpha_{2,i}}^2-\left(\frac{{\beta_{1,i}}}{\cos\hat{\Theta}_{L,i}+\tilde{\theta}_{L,i}\sin\hat{\Theta}_{L,i}}\right)^2\right],&& {B_{1,i}\leq\tilde{\theta}_{L,i}< B_{3,i}}\\
\frac{\hat{{X_i}}\hat{{Z_i}}(\sin\hat{\Theta}_{L,i}-\tilde{\theta}_{L,i}\cos\hat{\Theta}_{L,i})}{2a\hat{Y_i}
(\cos\hat{\Theta}_{L,i}+\tilde{\theta}_{L,i}\sin\hat{\Theta}_{L,i})^2}, &&
 {B_{3,i}\leq\tilde{\theta}_{L,i}< B_{2,i}}\\
 \frac{\hat{{X_i}}(\sin\hat{\Theta}_{L,i}-\tilde{\theta}_{L,i}\cos\hat{\Theta}_{L,i})}{8ab\hat{Y_i}}
\left[\left(\frac{{\beta_{2,i}}}{\cos\hat{\theta}_{L,i}+\Delta\theta_{L,i}\sin\hat{\theta}_{L,i}}\right)^2-{\alpha_{1,i}}^2\right], && {B_{2,i}\leq\tilde{\theta}_{L,i}\leq B_{4,i}}\\
0, && {\rm others.}
\end{array} \right.\nonumber\\\end{comment}
{Case 2}: When $\frac{{\beta_{2,i}}}{{\alpha_{2,i}}}<\frac{{\beta_{1,i}}}{{\alpha_{1,i}}}$, as we have \eqref{eeq40}, we could derive the approximation of $f_{\tilde{\Theta}_{L,i}}(\tilde{\theta}_{L,i})$, which is written as
\begin{align}\label{ee40}
f_{\tilde{\theta}_{L,i}}(\tilde{\theta}_{L,i})
&\approx\left\{
\begin{array}{rcl}
\frac{\hat{{X_i}}(\hat{V_i}-\tilde{\theta}_{L,i}\hat{{U_i}})}{8a_ib_i\hat{Y_i}}
\left[{\alpha_{2,i}}^2-\left(\frac{{\beta_{1,i}}}{\hat{{U_i}}+\tilde{\theta}_{L,i}\hat{V_i}}\right)^2\right],& & {B_{1,i}\leq\tilde{\theta}_{L,i}<B_{3,i}}\\
\frac{\hat{{X_i}}\hat{{Z_i}}(\hat{V_i}-\tilde{\theta}_{L,i}\hat{{U_i}})}
{2a\hat{Y_i}(\hat{{U_i}}+\tilde{\theta}_{L,i}\hat{V_i})^2}, & &
 {B_{3,i}\leq\tilde{\theta}_{L,i}<B_{2,i}}\\
 \frac{\hat{{X_i}}(\hat{V_i}-\tilde{\theta}_{L,i}\hat{{U_i}})}{8a_ib_i\hat{Y_i}}
\left[\left(\frac{{\beta_{2,i}}}{\hat{{U_i}}+\tilde{\theta}_{L,i}\hat{V_i}}\right)^2-{\alpha_{1,i}}^2\right], & & {B_{2,i}\leq\tilde{\theta}_{L,i}\leq B_{4,i}}\\
0, & & {\rm others.}
\end{array} \right.
\end{align}

\section{Variance of Angle Estimation Error} \label{Variance}
In this section, we aim to calculate the variance of $\tilde{\Phi}_{L,i}$ and $\tilde{\Theta}_{L,i}$ by using the PDF in the above section, which will be used for the 3D position estimation in the next section.
\subsection{Variance of $\tilde{\Phi}_{L,i}$}
In this subsection, we provide the variance expression of $\tilde{\Phi}_{L,i}$. Based on the PDF of $\tilde{\Phi}_{L,i}$ in \eqref{1e17}, the variance of $\tilde{\Phi}_{L,i}$ can be calculated as
\begin{align}\label{e17}\hspace{-1cm}
D(\tilde{\phi}_{L,i})&=E(\tilde{\phi}_{L,i}^2)-(E(\tilde{\phi}_{L,i}))^2\nonumber\\
&=\int^{\hat{\Phi}_{L,i}-a_{2,i}}_{\hat{\Phi}_{L,i}-a_{1,i}}\tilde{\phi}^2_{L,i}
f_{\tilde{\Phi}_{L,i}}(\tilde{\phi}_{L,i})d\tilde{\phi}_{L,i}
-\left(\int^{\hat{\Phi}_{L,i}-a_{2,i}}_{\hat{\Phi}_{L,i}-a_{1,i}}\tilde{\phi}_{L,i}f_{\tilde{\Phi}_{L,i}}(\tilde{\phi}_{L,i})d\tilde{\phi}_{L,i}\right)^2\nonumber\\
&= \frac{1}{2a_i}\left(\frac{\tilde{\phi}^3_{L,i}}{3}\cos\hat{\Phi}_{L,i}+\frac{\tilde{\phi}^4_{L,i}}{4}
\sin\hat{\Phi}_{L,i}\right)\bigg|^{\hat{\Phi}_{L,i}-a_{2,i}}_{\hat{\Phi}_{L,i}-a_{1,i}}\nonumber\\
&\quad-\frac{1}{4a_i^2}\left[\left(\frac{\tilde{\phi}^2_{L,i}}{2}\cos\hat{\Phi}_{L,i}+\frac{\tilde{\phi}^3_{L,i}}{3}
\sin\hat{\Phi}_{L,i}\right)\bigg|^{\hat{\Phi}_{L,i}-a_{2,i}}_{\hat{\Phi}_{L,i}-a_{1,i}}\right]^2,
\end{align}
where $f(x)\big|^{x_1}_{x_2}=f(x_1)-f(x_2)$ and $E(\tilde{\phi}_{L,i})$ denotes the expectation of $\tilde{\phi}_{L,i}$.
\subsection{Variance of $\tilde{\Theta}_{L,i}$}
In this subsection, we aim to derive the variance of $\tilde{\Theta}_{L,i}$. Different from the PDF of $\tilde{\Phi}_{L,i}$, the PDF of $\tilde{\Theta}_{L,i}$ is more complicated. Therefore, we need to analyze the variance according to the aforementioned two cases as follows.

Case 1: When $\frac{{\beta_{2,i}}}{{\alpha_{2,i}}}>\frac{{\beta_{1,i}}}{{\alpha_{1,i}}}$, based on the PDF of $\tilde{\theta}_{L,i}$ in \eqref{e40}, the variance of $\tilde{\Theta}_{L,i}$ is derived as
\begin{align}\label{e42}
D(\tilde{\theta}_{L,i})=
\underbrace{\int\tilde{\theta}^2_{L,i}f_{\tilde{\Theta}_{L,i}}(\tilde{\theta}_{L,i})d\tilde{\theta}_{L,i}}_{D_{1,i}}-
\bigg(\underbrace{\int\tilde{\theta}_{L,i} f_{\tilde{\Theta}_{L,i}}(\tilde{\theta}_{L,i})d\tilde{\theta}_{L,i}}_{D_{2,i}}\bigg)^2.
\end{align}

For $D_{1,i}$ in \eqref{e42}, we divide it into three different non-zero intervals that can be expressed as
\begin{align}\label{e43}
D_{1,i} = D_{11,i}+D_{12,i}+D_{13,i},
\end{align}
where $D_{11,i}$, $D_{12,i}$ and $D_{13,i}$ are the integral expressions in the intervals of $[B_{1,i},B_{2,i})$, $[B_{2,i},B_{3,i})$ and $[B_{3,i},B_{4,i}]$, respectively. The expressions of $D_{11,i}$, $D_{12,i}$ and $D_{13,i}$ are given in Appendix A.

For $D_{2,i}$ in \eqref{e42}, it is the expectation of $\tilde{\theta}_{L,i}$, which is denoted by $E(\tilde{\theta}_{L,i})$. It can be divided into three different non-zero intervals that are given by
\begin{align}\label{e46}
D_{2,i} = D_{21,i}+D_{22,i}+D_{23,i},
\end{align}
where $D_{21,i}$, $D_{22,i}$ and $D_{23,i}$ are the integral expressions in the intervals of $[B_{1,i},B_{2,i})$, $[B_{2,i},B_{3,i})$ and $[B_{3,i},B_{4,i}]$, respectively. The expressions of $D_{21,i}$, $D_{22,i}$ and $D_{23,i}$ are given in Appendix B.

Case 2: When $\frac{{\beta_{2,i}}}{{\alpha_{2,i}}}<\frac{{\beta_{1,i}}}{{\alpha_{1,i}}}$, according to the PDF of $\tilde{\Theta}_{L,i}$ in \eqref{ee40}, the variance of $\tilde{\Theta}_{L,i}$ can be calculated as
\begin{align}\label{e50}
D'(\tilde{\theta}_{L,i})=\underbrace{\int\tilde{\theta}^2_{L,i}f_{\tilde{\Theta}_{L,i}}(\tilde{\theta}_{L,i})d\tilde{\theta}_{L,i}}_{D'_{1,i}}
-\bigg(\underbrace{\int\tilde{\theta}_{L,i} f_{\tilde{\Theta}_{L,i}}(\tilde{\theta}_{L,i})d\tilde{\theta}_{L,i}}_{D'_{2,i}}\bigg)^2.
\end{align}

For $D'_{1,i}$, we divide it into three different non-zero intervals that are given by
\begin{align}\label{e51}
D'_{1,i} = D'_{11,i}+D'_{12,i}+D'_{13,i},
\end{align}
where $D'_{11,i}$, $D'_{12,i}$ and $D'_{13,i}$ are the integral expressions in the intervals of $[B_{1,i},B_{3,i})$, $[B_{3,i},B_{2,i})$ and $[B_{2,i},B_{4,i}]$. The expressions of $D'_{11,i}$, $D'_{12,i}$ and $D'_{13,i}$ are given in Appendix C.

For $D'_{2,i}$ in \eqref{e50}, it is the expectation of $\tilde{\theta}_{L,i}$, which is denoted by $E(\tilde{\theta}_{L,i})$. It is also divided into three different non-zero intervals that are given by
\begin{align}\label{e55}
D'_{2,i} =D'_{21,i}+D'_{22,i}+D'_{23,i},
\end{align}
where $D'_{21,i}$, $D'_{22,i}$ and $D'_{23,i}$ are the integral expressions in the intervals of $[B_{1,i},B_{3,i})$, $[B_{3,i},B_{2,i})$ and $[B_{2,i},B_{4,i}]$. The expressions of $D'_{21,i}$, $D'_{22,i}$ and $D'_{23,i}$ are given in Appendix D.

\begin{comment}
	conte, and the locations of anchor and M{U_i} are  ${\bf q}=[0,0,40]^T$ and ${\bf p}=[90,30,0]^T$, while the locations of three RISs are ${\bf s1}=[60,45,15]^T$,
	${\bf s2}=[50,50,5]^T$ and ${\bf s3}=[40,20,10]^T$nt... \eqref{1em18}
\end{comment}
\section{3D Position Estimation} \label{WLS}
Using the estimated AoA  in Section \ref{angle estimation algorithm} and the variance of the estimation error in Section \ref{Variance}, we aim to derive the expression of the estimation of the MU's 3D position at this section. First, we have the AoAs at the $i$th anchor given by
\begin{align}\label{bp1}
\hat{\Theta}_{L,i} &= \Theta_{L,i}+\tilde{\Theta}_{L,i}\nonumber\\
\hat{\Phi}_{L,i} &= \Phi_{L,i}+\tilde{\Phi}_{L,i},
\end{align}
where $\hat{\Theta}_{L,i}$ and $\hat{\Phi}_{L,i}$ denote the estimated AoAs at the $i$th anchor, ${\Theta}_{L,i}$ and ${\Phi}_{L,i}$ denote the true AoAs at the $i$th anchor, and $\tilde{\Theta}_{L,i}$ and $\tilde{\Phi}_{L,i}$ denote the estimation error at the $i$th anchor. For the sake of illustration, we collect all the estimated AoAs in the following vectors:
		\begin{align}\label{bp2}
			\hat{{\bm \Theta}}&={\bm \Theta}+\tilde{\bm \Theta},\nonumber\\
			\hat{\bm{\Phi}}&=\bm{\Phi}+\tilde{\bm \Phi},
		\end{align}
where
\begin{align}\label{bp3}
			\hat{{\bm \Theta}}&=[\hat{\Theta}_{L,1},\cdots,\hat{\Theta}_{L,I}],\quad
			\hat{\bm{\Phi}}=[\hat{\Phi}_{L,1},\cdots,\hat{\Phi}_{L,I}],\nonumber\\
{{\bm \Theta}}&=[{\Theta}_{L,1},\cdots,{\Theta}_{L,I}],\quad
			{\bm{\Phi}}=[{\Phi}_{L,1},\cdots,{\Phi}_{L,I}],\nonumber\\
\tilde{{\bm \Theta}}&=[\tilde{\Theta}_{L,1},\cdots,\tilde{\Theta}_{L,I}],\quad
			\tilde{\bm{\Phi}}=[\tilde{\Phi}_{L,1},\cdots,\tilde{\Phi}_{L,I}].
		\end{align}

In the existing works,  for tractability,  $\tilde{\bm \Theta}$ and $\tilde{\bm{\Phi}}$ are assumed to be the additive complex Gaussian noise with zero mean. However, according to the angle estimation error analysis in our previous section, the PDF of the estimation error $\tilde{\bm{\Phi}}$ should be modeled as \eqref{1e17}, while the PDF of the estimation error $\tilde{\bm \Theta}$ should be modeled as \eqref{e40} or \eqref{ee40}. Furthermore, the covariance matrices can be written as
		\begin{align} \label{bp4}
				{\bf Q}_{\Theta}=\textrm{diag}[\sigma^2_{\Theta_{L,1}},\cdots,\sigma^2_{\Theta_{L,I}}],\nonumber\\
                {\bf Q}_{\Phi}=\textrm{diag}[\sigma^2_{\Phi_{L,1}},\cdots,\sigma^2_{\Phi_{L,I}}],
\end{align}
where $\sigma^2_{\Theta_{L,i}}$ and $\sigma^2_{\Phi_{L,i}}$  denote the variance of $\tilde{\Theta}_{L,i}$ and $\tilde{\Phi}_{i}$, respectively. $\sigma^2_{\Phi_{L,i}}$  can be calculated according to \eqref{e17}, and $\sigma^2_{\Theta_{L,i}}$  can be calculated according to \eqref{e42} or \eqref{e50}.
\begin{comment}
Based on the variance derived at the last section, we derive the variance based weighted least square algorithm to localize the position of the MU with the AoA at the anchor.
\end{comment}

Accordingly, we propose to derive the closed-form expression of the MU's estimated 3D position. First, we can derive the pseudolinear equations \cite{Wu2} based on the estimated AoAs as follows:
		\begin{align} \label{bp5}
				\hat{\mathbf g}_{\Theta_{i}}^T\mathbf{s}_{i}-\hat{\mathbf{g}}_{\Theta_{i}}^T\mathbf{q}&\approx-\tilde\Theta_{L,i}d_{L,i}\cos\hat{\Phi}_{L,i},\nonumber\\
\hat{\mathbf g}_{\Phi_{i}}^T\mathbf{s}_{i}-\hat{\mathbf{g}}_{\Phi_{i}}^T\mathbf{q}&\approx-\tilde\Theta_{L,i}d_{L,i},
\end{align}
where $d_{L,i}$ denotes the distance between the $i$th anchor and  the MU, and we have
	\begin{align} \label{bp6}
				\hat{\mathbf g}_{\Theta_{i}}=[-\cos\hat{\Theta}_{L,i},\sin\hat{\Theta}_{L,i},0]^T,
\hat{\mathbf g}_{\Phi_{i}}=[-\sin\hat{\Phi}_{L,i}\sin\hat{\Theta}_{L,i},\sin\hat{\Phi}_{L,i}\cos\hat{\Theta}_{L,i},-\cos\hat{\Phi}_{L,i}]^T.
\end{align}
As a result, we can derive the following compact form of equations as
	\begin{align} \label{bp7}
				\hat{\mathbf h}-\hat{\mathbf G}\mathbf{q}={\bf Bz}
\end{align}
 where
 	\begin{align} \label{bp8}
				&\hat{\mathbf{h}}=[\mathbf{1}^T(\hat{\mathbf{G}}_{\Theta}\odot\mathbf{S})^T,\mathbf{1}^T(\hat{\mathbf{G}}_{\Phi}\odot\mathbf{S})^T]^T,
    \hat{ \mathbf{G}}=[\hat{\mathbf{G}}^T_{\Theta},\hat{\mathbf{G}}^T_{\Phi}]^T,
    {\bf B}= [{\bf B}_{1},{\bf B}_{2}]^T,
    {\bf z} = [\tilde{{\bm \Theta}}^T,\tilde{{\bm \Phi}}^T]^T,\nonumber\\
     &\hat{\mathbf{G}}_{\Theta}= [\hat{\mathbf g}_{\Theta_{1}},\cdots,\hat{\mathbf g}_{\Theta_{I}}]^T,
    \hat{\mathbf{G}}_{\Phi} = [\hat{\mathbf g}_{\Phi_{1}},\cdots,\hat{\mathbf g}_{\Phi_{I}}]^T,
    \mathbf{S}=[\mathbf{s}_{{1}},\cdots,\mathbf{s}_{{I}}]^T,
    {\bf B}_{1}= [{\bf B}_{\Theta},{\bf O}]^T,
    {\bf B}_{2}= [{\bf O},{\bf B}_{\Phi}]^T,\nonumber\\
    &\mathbf{B}_{\Theta}  =-\textrm{diag}[d_{L,1}\cos\hat{\Phi}_{L,1},\cdots,d_{L,I}\cos\hat{\Phi}_{L,I}]^T,
    \mathbf{B}_{\Phi}  =-\textrm{diag}[d_{L,1},\cdots,d_{L,I}]^T.
\end{align}

Based on \eqref{bp7}, we can apply the WLS algorithm \cite{Wu2} to derive the closed-form expression of the MU's estimated position, which is written as:
	\begin{align} \label{bp8}
				\hat{\bf q}=(\hat{\mathbf{G}}^T\mathbf{W}\hat{\mathbf{G}})^{-1}\hat{\mathbf{G}}^T\mathbf{W}\hat{\mathbf{h}},
\end{align}
where  $\mathbf{W} =  {\bf B} {\bf Q} {\bf B}^T$ and $ \mathbf{Q} = \textrm{diag}[	\mathbf{Q}_{\Theta},\mathbf{Q}_{\Phi}]$. The details of the derivation can be found in \cite{Wu2}.

\begin{figure*}
					\setlength{\abovecaptionskip}{-5pt}
					\setlength{\belowcaptionskip}{-15pt}
					\centering
					\begin{minipage}[t]{0.46\linewidth}
						\centering
						\includegraphics[width= 1\textwidth]{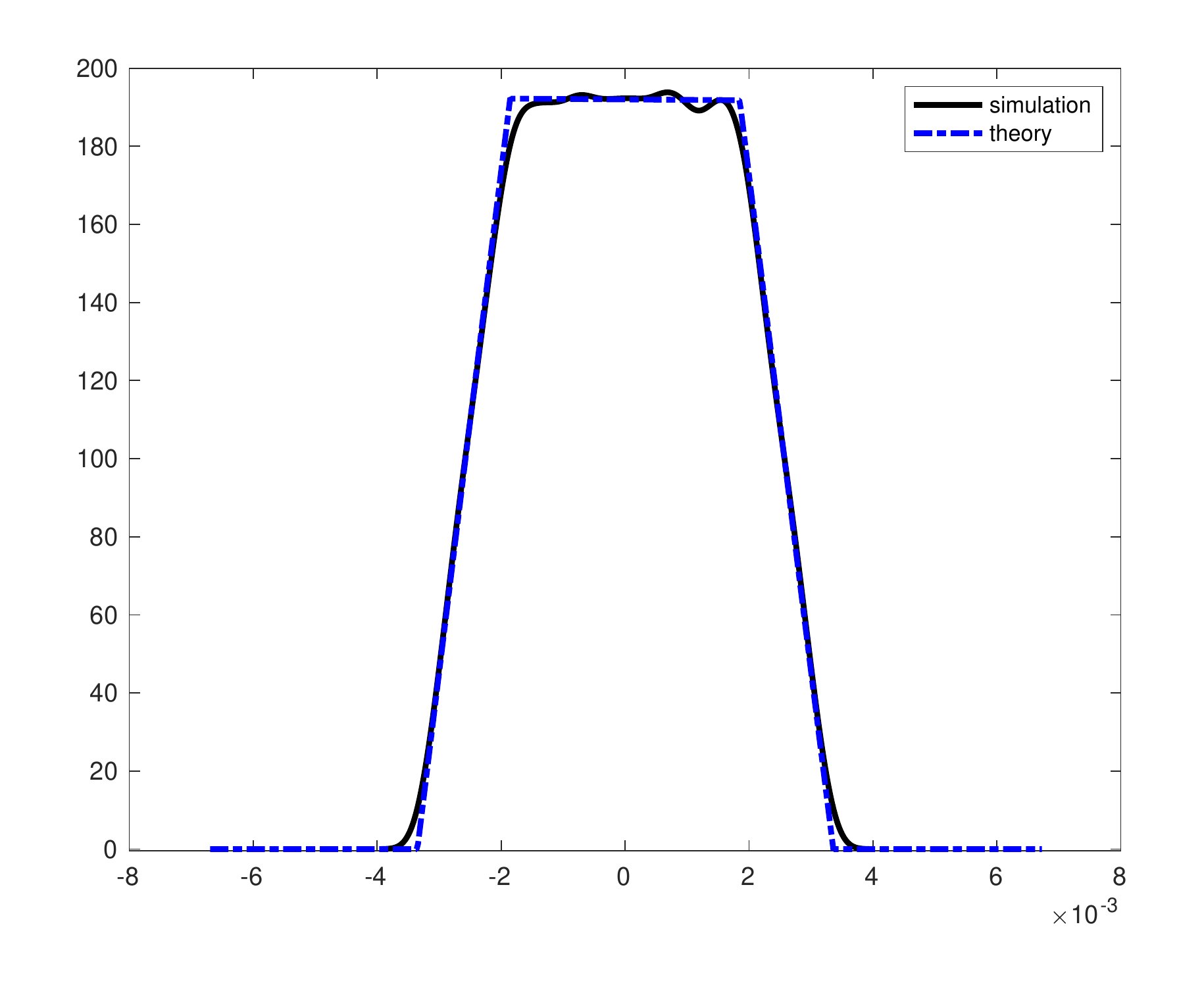}
						\DeclareGraphicsExtensions.
						\captionsetup{font={small}}
						\caption{PDF of $\tilde{\Theta}_{L,i}$.}
						\label{pdf_theta_ar}
					\end{minipage}
					\begin{minipage}[t]{0.45\linewidth}
						\centering
						\includegraphics[width= 1\textwidth]{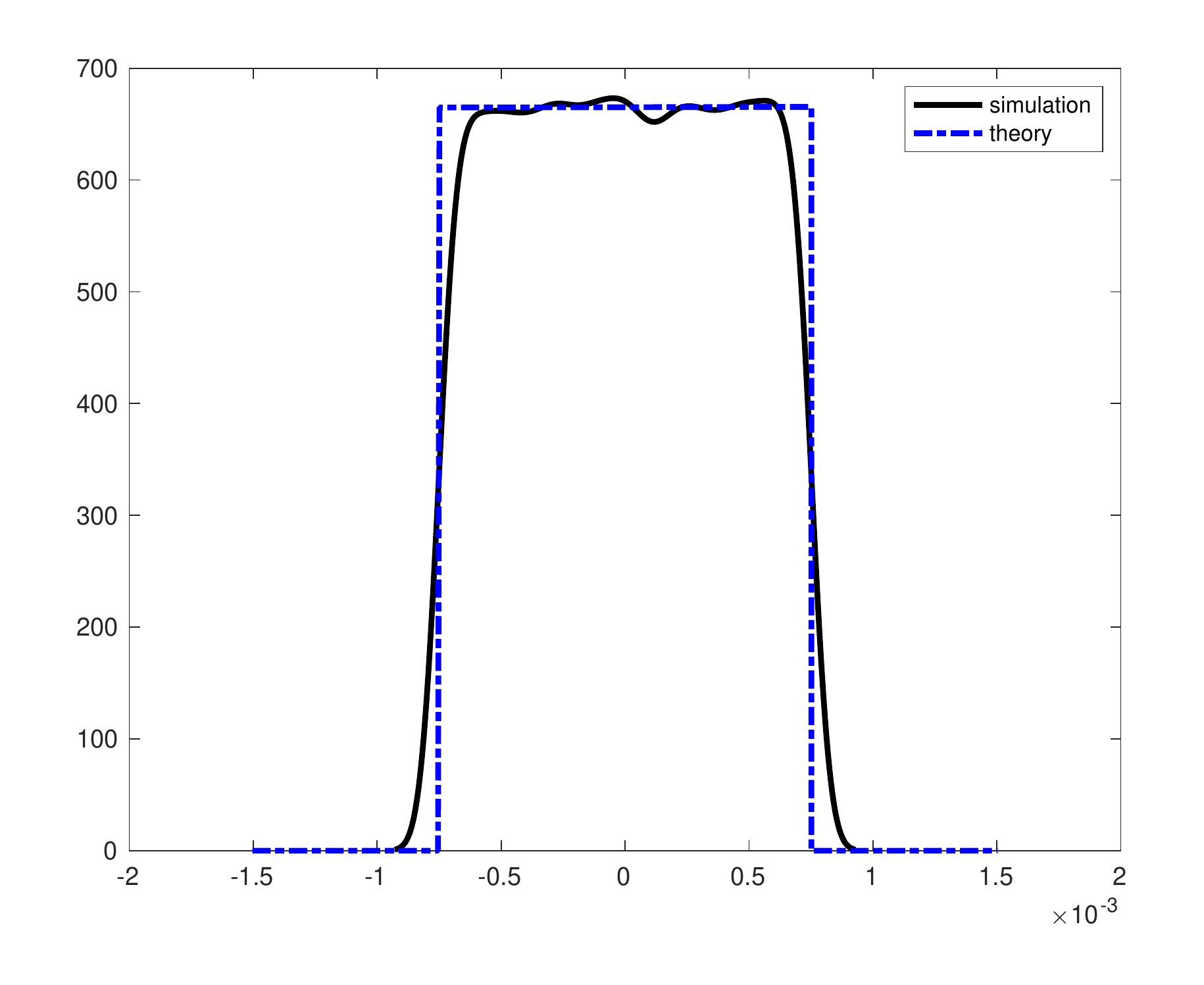}
						\DeclareGraphicsExtensions.
						\captionsetup{font={small}}
						\caption{ PDF of $\tilde{\Phi}_{L,i}$.}
						\label{pdf_phi_ar}
					\end{minipage}
				\end{figure*}

\begin{figure*}
					\setlength{\abovecaptionskip}{-5pt}
					\setlength{\belowcaptionskip}{-15pt}
					\centering
					\begin{minipage}[t]{0.46\linewidth}
						\centering
						\includegraphics[width= 1\textwidth]{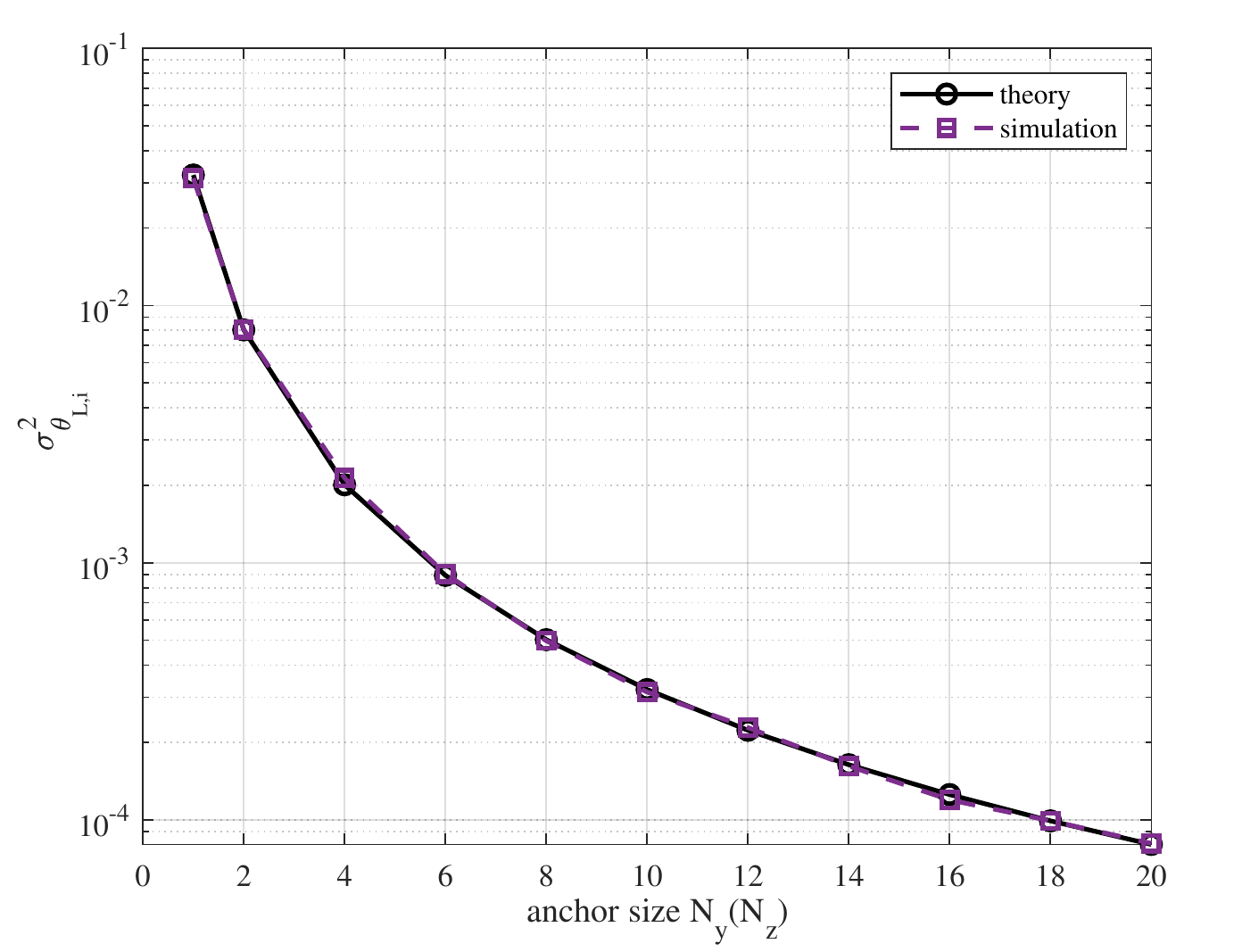}
						\DeclareGraphicsExtensions.
						\captionsetup{font={small}}
						\caption{Variance of $\tilde{\Theta}_{L,i}$ versus anchor size.}
						\label{var_theta}
					\end{minipage}
					\begin{minipage}[t]{0.44\linewidth}
						\centering
						\includegraphics[width= 1\textwidth]{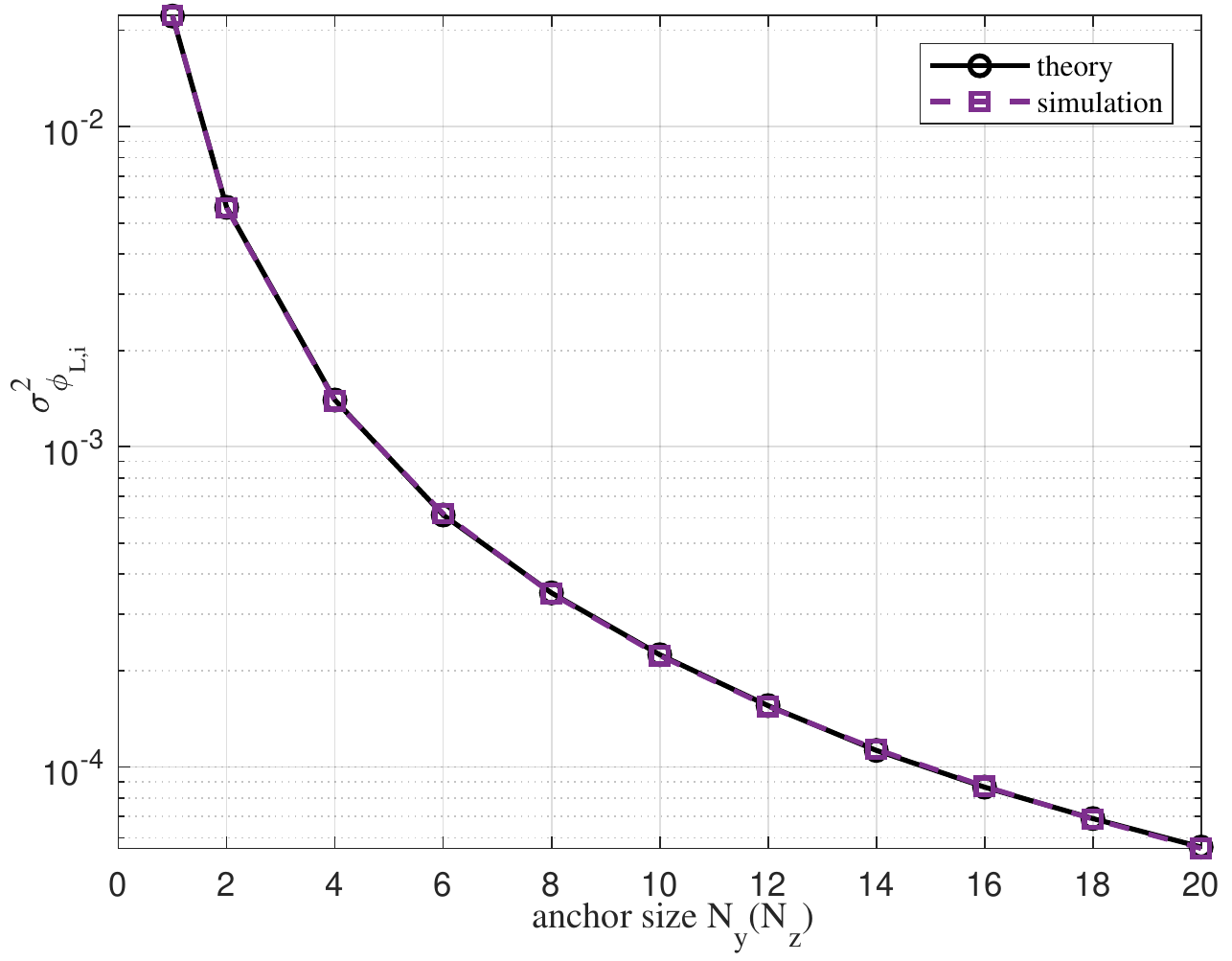}
						\DeclareGraphicsExtensions.
						\captionsetup{font={small}}
						\caption{Variance of $\tilde{\Phi}_{L,i}$ versus anchor size.}
						\label{var_phi}
					\end{minipage}
				\end{figure*}
\section{Simulation Results} \label{result}
This section presents simulation results to validate the accuracy of our derivations and approximations. In our simulation, we consider a mmWave  multiple input single  output (MISO) channel from the MU to the anchors.  Moreover, the MU, the anchors are assumed to be placed in a 3D  area. The locations of four anchors are ${\bf s}_1=[2,20,3]^T$, ${\bf s}_2=[-12,-16,58]^T$, ${\bf s}_3=[-10,-6,-8]^T$ and ${\bf s}_4=[10,6,-20]^T$.   It is assumed that the inter-antenna spacing of UPA at the anchors is $d_r = \lambda_c/2$. The following results are obtained by averaging over 10,000 random estimation error realizations. Unless otherwise stated, we assume $S=S_{1}\times S_{2}=64\times64$  for rotation angle search grids, and the SNR is assumed to be $10$ dB.  The positioning accuracy is assessed in terms of the mean-square error (MSE).

 Fig. \ref{pdf_theta_ar} and Fig. \ref{pdf_phi_ar} illustrate the PDF of estimation errors. It is assumed that the anchor size is $N_y=N_z=16$. It can be observed from Fig. \ref{pdf_theta_ar} and Fig. \ref{pdf_phi_ar} that our derived results match well with the simulation results, which verify the accuracy of our derived results and confirm that the angle estimation error is non-Gaussian.

Fig. \ref{var_theta} and Fig. \ref{var_phi} display the variance $\sigma^2_{\theta_{L,i}}$ of estimation error $\tilde{\Theta}_{L,i}$ and the variance $\sigma^2_{\phi_{L,i}}$ of estimation error $\tilde{\Phi}_{L,i}$  as the functions of anchor size $N_y(N_z)$, respectively.  Fig. \ref{var_theta} and Fig. \ref{var_phi} display the variance $\sigma^2_{\theta_{L,i}}$, which show that the theoretical results coincide with the simulation results, which validates the correctness of the derived results. Moreover, it is observed that the variances of $\tilde{\Theta}_{L,i}$ and $\tilde{\Phi}_{L,i}$ decrease with the anchor size, which means that increasing the number of antennas could improve the estimation accuracy.

\begin{figure*}
					\setlength{\abovecaptionskip}{-5pt}
					\setlength{\belowcaptionskip}{-15pt}
					\centering
					\begin{minipage}[t]{0.46\linewidth}
						\centering
						\includegraphics[width= 1\textwidth]{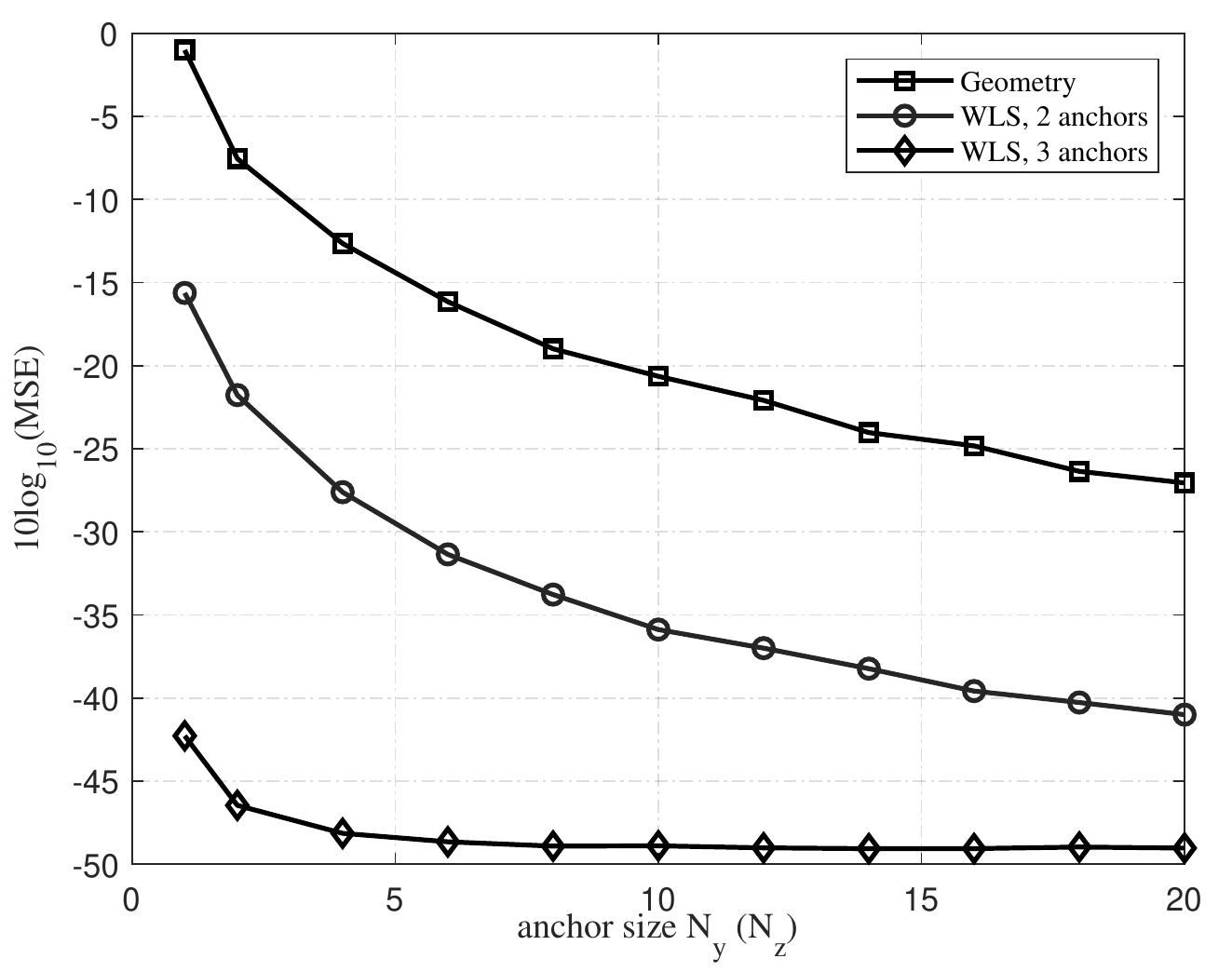}
						\DeclareGraphicsExtensions.
						\captionsetup{font={small}}
						\caption{ Comparison of the proposed framework and geometry algorithm.}
						\label{anchor_algorithm}
					\end{minipage}
					\begin{minipage}[t]{0.45\linewidth}
						\centering
						\includegraphics[width= 1\textwidth]{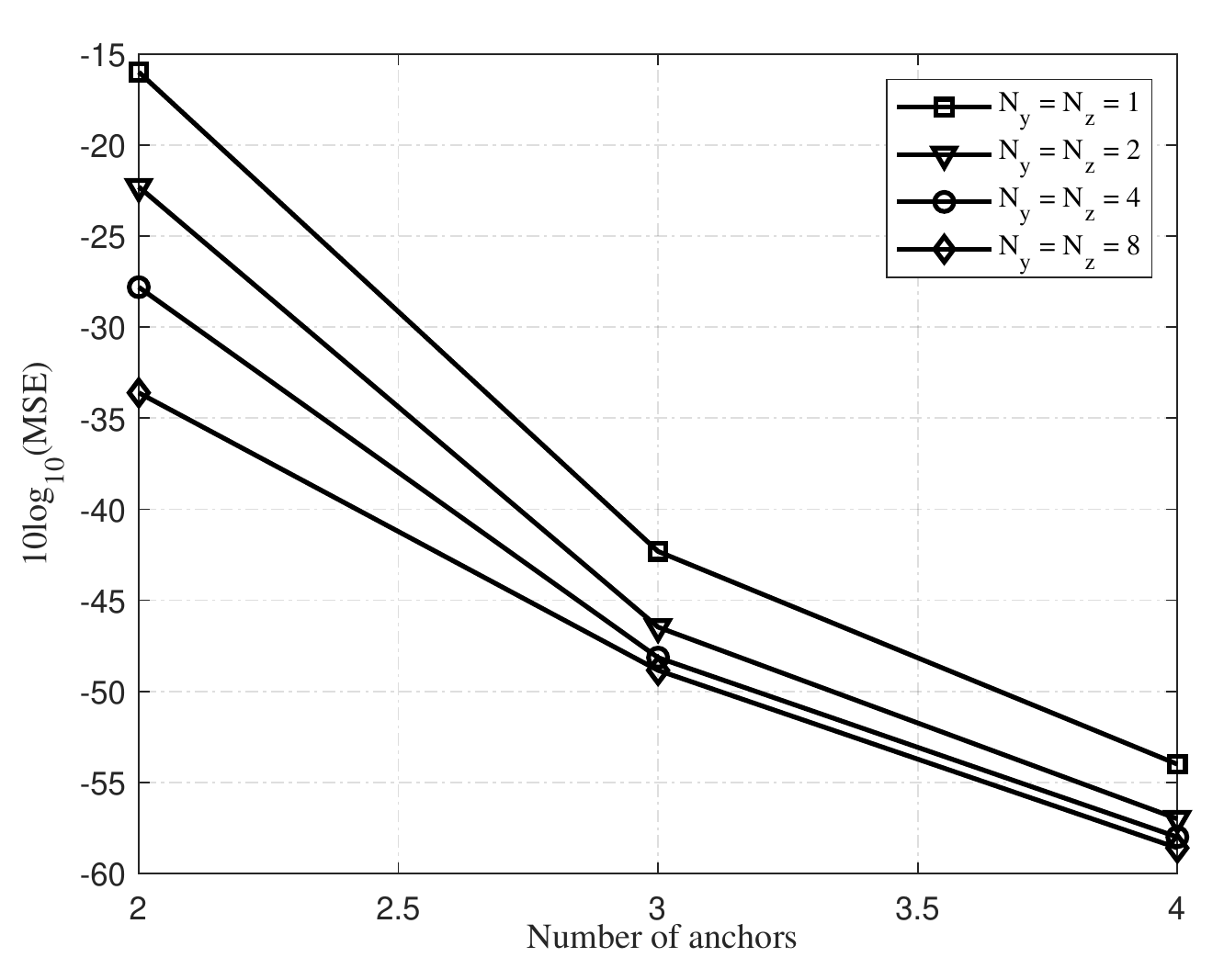}
						\DeclareGraphicsExtensions.
						\captionsetup{font={small}}
						\caption{ MSE of the proposed framework versus anchors number.}
						\label{achor_number}
					\end{minipage}
				\end{figure*}

Fig. \ref{anchor_algorithm} compares the MSE of the proposed framework aided by 2 anchors with that aided by 3 anchors when the size of anchors increases from $N_y=N_z= 1$  to $N_y=N_z= 20$. As shown in the figure, the MSE decreases  with the anchor size as expected. Furthermore,  it is shown that the MSE of 2 anchors is larger than that of 3 anchors.  It implies that increasing the number of anchors can significantly improve the positioning accuracy of the proposed framework.

In \cite{Henk}, the position of the MU is derived by using the geometry relationship between the estimated AOA and the 3D position, which is denoted as the geometry algorithm in this paper. Fig. \ref{anchor_algorithm} also presents the positioning performance comparison of the proposed framework with the geometry algorithm. It is seen from Fig. \ref{anchor_algorithm} that the proposed framework outperforms the geometry algorithm even with only 2 anchors.

Fig. \ref{achor_number} illustrates the impact of the number of anchors and the size of the anchors on the performance of the proposed framework. To be specific, we present the comparison of the proposed framework with $N_y=N_z=1$, $N_y=N_z=2$, $N_y=N_z=4$, and $N_y=N_z=8$. As it can be seen from Fig. \ref{achor_number}, MSE decreases with the number of anchors. This implies that the positioning accuracy is better with a larger number of anchors. Furthermore, as shown in Fig. \ref{achor_number}, there is a large gap between the curve of $N_y=N_z=1$ and the curve of  $N_y=N_z=2$, while the curve of $N_y=N_z=8$  is only a little better than the curve of $N_y=N_y=4$. This implies that the impact of the anchor size becomes small when $N_y (N_z)$ is larger than $4$.

\section{Conclusion}\label{Con}
   In this paper, we designed a comprehensive framework to analyze the angle estimation error and design the 3D positioning algorithm for the mmWave system. First, we estimated the AoAs at the anchors by applying the 2D-DFT algorithm.  Based on the property of the 2D-DFT algorithm,  the angle estimation error was analyzed in terms of PDF.  We then simplified the intricate geometric expression of the error PDF by employing the first-order linear approximation of triangle functions.   We also derived the variance expression of the error by using the error PDF and theoretically derived the variance from the error PDF.  Finally,  we applied the WLS algorithm to estimate the 3D position of the MU by using the estimated AoAs and the obtained non-Gaussian variance. Extensive simulation results confirmed that the derived  angle estimation error is non-Gaussian, and also demonstrated the superiority of the proposed framework.
 \begin{appendices}
 \section{Derivation of $D_{1,i}$ }
Firstly, we derive the expression of $D_{11,i}$ as
\begin{align}\label{e44}
D_{11,i} 
&=\frac{\hat{{X_i}}{\alpha_{2,i}}^2}{8a_ib_i\hat{Y_i}}\int_{B_{1,i}}^{B_{2,i}}(\hat{V_i}\tilde{\theta}^2_{L,i}
-\hat{U_i}\tilde{\theta}^3_{L,i}) d\tilde{\theta}_{L,i}-
\frac{\hat{{X_i}}{\beta_{1,i}}^2}{8a_ib_i\hat{Y_i}}\int_{B_{1,i}}^{B_{2,i}}\frac{\hat{V_i}\tilde{\theta}^2_{L,i}-\hat{U_i}\tilde{\theta}^3_{L,i}}
{(\hat{U_i}+\hat{V_i}\tilde{\theta}_{L,i})^2}d\tilde{\theta}_{L,i}\nonumber\\
&=\frac{\hat{{X_i}}{\alpha_{2,i}}^2}{8a_ib_i\hat{Y_i}}\int_{B_{1,i}}^{B_{2,i}}(\hat{V_i}\tilde{\theta}^2_{L,i}
-\hat{U_i}\tilde{\theta}^3_{L,i})d\tilde{\theta}_{L,i}\nonumber\\
&\quad-\bigg[\frac{\hat{{X_i}}{\beta_{1,i}}^2}{8a_ib_i\hat{Y_i}\hat{U_i}}\bigg(\int_{0}^{B_{2,i}}\frac{\frac{\hat{V_i}}{\hat{U_i}}\tilde{\theta}^2_{L,i}}
{(1+\frac{\hat{V_i}\tilde{\theta}_{L,i}}{\hat{U_i}})^2}d\tilde{\theta}_{L,i}-\int_{0}^{B_{2,i}}\frac{\tilde{\theta}^3_{L,i}}
{(1+\frac{\hat{V_i}\tilde{\theta}_{L,i}}{\hat{U_i}})^2}d\tilde{\theta}_{L,i}
\bigg)\nonumber\\
&=\frac{\hat{{X_i}}{\alpha_{2,i}}^2}{8a_ib_i\hat{Y_i}}\bigg(\frac{\hat{U_i}\tilde{\theta}^4_{L,i}}{4}-
\frac{\hat{V_i}\tilde{\theta}^3_{L,i}}{3}\bigg)\bigg|_{B_{1,i}}^{B_{2,i}}-\nonumber\\
&\quad\bigg[\bigg(-\frac{\hat{{X_i}}{\beta_{1,i}}^2}{8a_ib_i\hat{Y_i}\hat{U_i}}\cdot\frac{B_{2,i}^4}{4} \cdot\ _2F_1\bigg(2,4;5;-\frac{\hat{V_i}}{\hat{U_i}}B_{2,i}\bigg)
+\frac{\hat{{X_i}}{\beta_{1,i}}^2\hat{V_i}}{8a_ib_i\hat{Y_i}\hat{U_i}^2}\cdot\frac{B_{2,i}^3}{3} \cdot\ _2F_1\bigg(2,3;4;-\frac{\hat{V_i}}{\hat{U_i}}B_{2,i}\bigg)\bigg)\nonumber\\
&\quad-\bigg(-\frac{\hat{{X_i}}{\beta_{1,i}}^2}{8a_ib_i\hat{Y_i}\hat{U_i}}\cdot\frac{B_{1,i}^4}{4} \cdot\ _2F_1\bigg(2,4;5;-\frac{\hat{V_i}}{\hat{U_i}}B_{1,i}\bigg)
+\frac{\hat{{X_i}}{\beta_{1,i}}^2\hat{V_i}}{8a_ib_i\hat{Y_i}\hat{U_i}^2}\cdot\frac{B_{1,i}^3}{3} \cdot\ _2F_1\bigg(2,3;4;-\frac{\hat{V_i}}{\hat{U_i}}B_{1,i}\bigg)\bigg)\bigg],
\end{align}
where $\int^u_0 \frac{x^{\mu-1}dx}{(1+\beta x)^v}=_2F_1\bigg(\nu,\mu;1+\mu;-\beta u\bigg)$ is a generalized hypergeometric series \cite{Kemp}.

Then, $D_{12,i}$ in \eqref{e43}  is derived as
\begin{align}\label{e45}
D_{12,i}= \int_{B_{2,i}}^{B_{3,i}}\frac{\hat{{X_i}}(\hat{V_i}-\tilde{\theta}_{L,i}\hat{U_i})}{2b}\tilde{\theta}^2_{L,i}d\tilde{\theta}_{L,i}
=\frac{\hat{{X_i}}}{2b_i}\bigg(\frac{\hat{V_i}\tilde{\theta}^3_{L,i}}{3}-\frac{\hat{U_i}\tilde{\theta}^4_{L,i}}{4}\bigg)\bigg|_{B_{2,i}}^{B_{3,i}}.
\end{align}
Finally, $D_{13,i}$ in \eqref{e43} is given by
\begin{align}\label{ee45}
&D_{13,i}
=\bigg[\bigg(-\frac{\hat{{X_i}}{\beta_{2,i}}^2}{8a_ib_i\hat{Y_i}\hat{U_i}}\cdot\frac{B_{4,i}^4}{4} \cdot\ _2F_1\bigg(2,4;5;-\frac{\hat{V_i}}{\hat{U_i}}B_{4,i}\bigg)
+\frac{\hat{{X_i}}{\beta_{2,i}}^2\hat{V_i}}{8a_ib_i\hat{Y_i}\hat{U_i}^2}\cdot\frac{B_{4,i}^3}{3} \cdot\ _2F_1\bigg(2,3;4;-\frac{\hat{V_i}}{\hat{U_i}}B_{4,i}\bigg)\bigg)\nonumber\\
&-\bigg(-\frac{\hat{{X_i}}{\beta_{2,i}}^2}{8a_ib_i\hat{Y_i}\hat{U_i}}\cdot\frac{B_{3,i}^4}{4} \cdot\ _2F_1\bigg(2,4;5;-\frac{\hat{V_i}}{\hat{U_i}}B_{3,i}\bigg)
+\frac{\hat{{X_i}}{\beta_{2,i}}^2\hat{V_i}}{8a_ib_i\hat{Y_i}\hat{U_i}^2}\cdot\frac{B_{3,i}^3}{3} \cdot\ _2F_1\bigg(2,3;4;-\frac{\hat{V_i}}{\hat{U_i}}B_{3,i}\bigg)\bigg)\bigg]\nonumber\\
&-\frac{\hat{{X_i}}{\alpha_{1,i}}^2}{8a_ib_i\hat{Y_i}}\bigg(-\frac{\hat{U_i}\tilde{\theta}^4_{L,i}}{4}+
\frac{\hat{V_i}\tilde{\theta}^3_{L,i}}{3}\bigg)\bigg|_{B_{3,i}}^{B_{4,i}}.
\end{align}
\section{Derivation of $D_{2,i}$ }
Firstly, $D_{21,i}$ in \eqref{e46} is derived as
\begin{align}\label{e47}
D_{21,i}
&=\frac{\hat{{X_i}}{\alpha_{2,i}}^2}{8a_ib_i\hat{Y_i}}\bigg(-\frac{\hat{U_i}\tilde{\theta}^3_{L,i}}{3}+
\frac{\hat{V_i}\tilde{\theta}^2_{L,i}}{2}\bigg)\bigg|_{B_{1,i}}^{B_{2,i}}\nonumber\\
&-\bigg[\bigg(-\frac{\hat{{X_i}}{\beta_{1,i}}^2}{8a_ib_i\hat{Y_i}\hat{U_i}}\cdot\frac{B_{2,i}^3}{3} \cdot\ _2F_1\bigg(2,3;4;-\frac{\hat{V_i}}{\hat{U_i}}B_{2,i}\bigg)
+\frac{\hat{{X_i}}{\beta_{1,i}}^2\hat{V_i}}{8a_ib_i\hat{Y_i}\hat{U_i}^2}\cdot\frac{B_{2,i}^2}{2} \cdot\ _2F_1\bigg(2,2;3;-\frac{\hat{V_i}}{\hat{U_i}}B_{2,i}\bigg)\bigg)\nonumber\\
&-\bigg(-\frac{\hat{{X_i}}{\beta_{1,i}}^2}{8a_ib_i\hat{Y_i}\hat{U_i}}\cdot\frac{B_{1,i}^3}{3} \cdot\ _2F_1\bigg(2,3;4;-\frac{\hat{V_i}}{\hat{U_i}}B_{1,i}\bigg)
+\frac{\hat{{X_i}}{\beta_{1,i}}^2\hat{V_i}}{8a_ib_i\hat{Y_i}\hat{U_i}^2}\cdot\frac{B_{1,i}^2}{2} \cdot\ _2F_1\bigg(2,2;3;-\frac{\hat{V_i}}{\hat{U_i}}B_{1,i}\bigg)\bigg)\bigg].
\end{align}

Then, $D_{22,i}$ in \eqref{e46} is written as
\begin{align}\label{e48}
D_{22,i} = \int_{B_{2,i}}^{B_{3,i}}\frac{\hat{{X_i}}(\hat{V_i}-\tilde{\theta}_{L,i}\hat{U_i})}{2b_i}\tilde{\theta}_{L,i}d\tilde{\theta}_{L,i}
=\frac{\hat{{X_i}}}{2b_i}\bigg(-\frac{\hat{U_i}\tilde{\theta}^3_{L,i}}{3}+\frac{\hat{V_i}\tilde{\theta}^2_{L,i}}{2}\bigg)\bigg|_{B_{2,i}}^{B_{3,i}}.
\end{align}

Finally, $D_{23,i}$ in \eqref{e46} could be formulated as
\begin{align}\label{e49}
D_{23,i}
&=\bigg[\bigg(-\frac{\hat{{X_i}}{\beta_{2,i}}^2}{8a_ib_i\hat{Y_i}\hat{U_i}}\cdot\frac{B_{4,i}^3}{3} \cdot\ _2F_1\bigg(2,3;4;-\frac{\hat{V_i}}{\hat{U_i}}B_{4,i}\bigg)
+\frac{\hat{{X_i}}{\beta_{2,i}}^2\hat{V_i}}{8a_ib_i\hat{Y_i}\hat{U_i}^2}\cdot\frac{B_{4,i}^2}{2} \cdot\ _2F_1\bigg(2,2;3;-\frac{\hat{V_i}}{\hat{U_i}}B_{4,i}\bigg)\bigg)\nonumber\\
&-\bigg(-\frac{\hat{{X_i}}{\beta_{2,i}}^2}{8a_ib_i\hat{Y_i}\hat{U_i}}\cdot\frac{B_{3,i}^3}{3} \cdot\ _2F_1\bigg(2,3;4;-\frac{\hat{V_i}}{\hat{U_i}}B_{3,i}\bigg)
+\frac{\hat{{X_i}}{\beta_{2,i}}^2\hat{V_i}}{8a_ib_i\hat{Y_i}\hat{U_i}^2}\cdot\frac{B_{3,i}^2}{2} \cdot\ _2F_1\bigg(2,2;3;-\frac{\hat{V_i}}{\hat{U_i}}B_{3,i}\bigg)\bigg)\bigg]\nonumber\\
&-\frac{\hat{{X_i}}{\alpha_{1,i}}^2}{8a_ib_i\hat{Y_i}}\bigg(-\frac{\hat{U_i}\tilde{\theta}^3_{L,i}}{3}+
\frac{\hat{V_i}\tilde{\theta}^2_{L,i}}{2}\bigg)\bigg|_{B_{3,i}}^{B_{4,i}}.
\end{align}
\section{Derivation of $D'_{1,i}$ }
As for $D'_{11,i}$ in \eqref{e51}, it could be expressed as
\begin{align}\label{e52}
D'_{11,i}
&=\frac{\hat{{X_i}}{\alpha_{2,i}}^2}{8a_ib_i\hat{Y_i}}\bigg(-\frac{\hat{U_i}\tilde{\theta}^4_{L,i}}{4}+
\frac{\hat{V_i}\tilde{\theta}^3_{L,i}}{3}\bigg)\bigg|_{B_{1,i}}^{B_{3,i}}-\nonumber\\
&\bigg[\bigg(-\frac{\hat{{X_i}}{\beta_{1,i}}^2}{8a_ib_i\hat{Y_i}\hat{U_i}}\cdot\frac{B_{3,i}^4}{4} \cdot\ _2F_1\bigg(2,4;5;-\frac{\hat{V_i}}{\hat{U_i}}B_{3,i}\bigg)
+\frac{\hat{{X_i}}{\beta_{1,i}}^2\hat{V_i}}{8a_ib_i\hat{Y_i}\hat{U_i}^2}\cdot\frac{B_{3,i}^3}{3} \cdot\ _2F_1\bigg(2,3;4;-\frac{\hat{V_i}}{\hat{U_i}}B_{3,i}\bigg)\bigg)\nonumber\\
&-\bigg(-\frac{\hat{{X_i}}{\beta_{1,i}}^2}{8a_ib_i\hat{Y_i}\hat{U_i}}\cdot\frac{B_{1,i}^4}{4} \cdot\ _2F_1\bigg(2,4;5;-\frac{\hat{V_i}}{\hat{U_i}}B_{1,i}\bigg)
+\frac{\hat{{X_i}}{\beta_{1,i}}^2\hat{V_i}}{8a_ib_i\hat{Y_i}\hat{U_i}^2}\cdot\frac{B_{1,i}^3}{3} \cdot\ _2F_1\bigg(2,3;4;-\frac{\hat{V_i}}{\hat{U_i}}B_{1,i}\bigg)\bigg)\bigg].
\end{align}
Then, $D'_{12,i}$ in \eqref{e51} is given by
\begin{align}\label{e53}
D'_{12,i}
&=\bigg(-\frac{\hat{{X_i}}\hat{{Z_i}}}{2a\hat{Y_i}\hat{U_i}}\cdot\frac{B_{2,i}^4}{4} \cdot\ _2F_1\bigg(2,4;5;-\frac{\hat{V_i}}{\hat{U_i}}B_{2,i}\bigg)
+\frac{\hat{{X_i}}\hat{{Z_i}}\hat{V_i}}{2a\hat{Y_i}\hat{U_i}^2}\cdot\frac{B_{2,i}^3}{3} \cdot\ _2F_1\bigg(2,3;4;-\frac{\hat{V_i}}{\hat{U_i}}B_{2,i}\bigg)\bigg)\nonumber\\
&-\bigg(-\frac{\hat{{X_i}}\hat{{Z_i}}}{2a\hat{Y_i}\hat{U_i}}\cdot\frac{B_{3,i}^4}{4} \cdot\ _2F_1\bigg(2,4;5;-\frac{\hat{V_i}}{\hat{U_i}}B_{3,i}\bigg)
+\frac{\hat{{X_i}}\hat{{Z_i}}\hat{V_i}}{2a\hat{Y_i}\hat{U_i}^2}\cdot\frac{B_{3,i}^3}{3} \cdot\ _2F_1\bigg(2,3;4;-\frac{\hat{V_i}}{\hat{U_i}}B_{3,i}\bigg)\bigg).
\end{align}
Finally, $D'_{13,i}$ in \eqref{e51} is derived as
\begin{align}\label{e54}
D'_{13,i}
&=\bigg[\bigg(-\frac{\hat{{X_i}}{\beta_{2,i}}^2}{8a_ib_i\hat{Y_i}\hat{U_i}}\cdot\frac{B_{4,i}^4}{4} \cdot\ _2F_1\bigg(2,4;5;-\frac{\hat{V_i}}{\hat{U_i}}B_{4,i}\bigg)
+\frac{\hat{{X_i}}{\beta_{2,i}}^2\hat{V_i}}{8a_ib_i\hat{Y_i}\hat{U_i}^2}\cdot\frac{B_{4,i}^3}{3} \cdot\ _2F_1\bigg(2,3;4;-\frac{\hat{V_i}}{\hat{U_i}}B_{4,i}\bigg)\bigg)\nonumber\\
&-\bigg(-\frac{\hat{{X_i}}{\beta_{2,i}}^2}{8a_ib_i\hat{Y_i}\hat{U_i}}\cdot\frac{B_{2,i}^4}{4} \cdot\ _2F_1\bigg(2,4;5;-\frac{\hat{V_i}}{\hat{U_i}}B_{2,i}\bigg)
+\frac{\hat{{X_i}}{\beta_{2,i}}^2\hat{V_i}}{8a_ib_i\hat{Y_i}\hat{U_i}^2}\cdot\frac{B_{2,i}^3}{3} \cdot\ _2F_1\bigg(2,3;4;-\frac{\hat{V_i}}{\hat{U_i}}B_{2,i}\bigg)\bigg)\bigg]\nonumber\\
&-\frac{\hat{{X_i}}{\alpha_{1,i}}^2}{8a_ib_i\hat{Y_i}}\bigg(-\frac{\hat{U_i}\tilde{\theta}^4_{L,i}}{4}+
\frac{\hat{V_i}\tilde{\theta}^3_{L,i}}{3}\bigg)\bigg|_{B_{2,i}}^{B_{4,i}}.
\end{align}
\section{Derivation of $D'_{2,i}$ }
Firstly, $D'_{21,i}$ in \eqref{e55} is formulated as
\begin{align}\label{e56}
D'_{21,i}
&=\frac{\hat{{X_i}}{\alpha_{2,i}}^2}{8a_ib_i\hat{Y_i}}\bigg(-\frac{\hat{U_i}\tilde{\theta}^3_{L,i}}{3}+
\frac{\hat{V_i}\tilde{\theta}^2_{L,i}}{2}\bigg)\bigg|_{B_{1,i}}^{B_{3,i}}-\nonumber\\
&\quad\bigg[\bigg(-\frac{\hat{{X_i}}{\beta_{1,i}}^2}{8a_ib_i\hat{Y_i}\hat{U_i}}\cdot\frac{B_{3,i}^3}{3} \cdot\ _2F_1\bigg(2,3;4;-\frac{\hat{V_i}}{\hat{U_i}}B_{3,i}\bigg)
+\frac{\hat{{X_i}}{\beta_{1,i}}^2\hat{V_i}}{8a_ib_i\hat{Y_i}\hat{U_i}^2}\cdot\frac{B_{3,i}^2}{2} \cdot\ _2F_1\bigg(2,2;3;-\frac{\hat{V_i}}{\hat{U_i}}B_{3,i}\bigg)\bigg)\nonumber\\
&\quad-\bigg(-\frac{\hat{{X_i}}{\beta_{1,i}}^2}{8a_ib_i\hat{Y_i}\hat{U_i}}\cdot\frac{B_{1,i}^3}{3} \cdot\ _2F_1\bigg(2,3;4;-\frac{\hat{V_i}}{\hat{U_i}}B_{1,i}\bigg)
+\frac{\hat{{X_i}}{\beta_{1,i}}^2\hat{V_i}}{8a_ib_i\hat{Y_i}\hat{U_i}^2}\cdot\frac{B_{1,i}^2}{2} \cdot\ _2F_1(2,2;3;-\frac{\hat{V_i}}{\hat{U_i}}B_{1,i})\bigg)\bigg].
\end{align}

Then, $D'_{22,i}$ in \eqref{e55} is written as
\begin{align}\label{e57}
D'_{22,i}
&=\bigg(-\frac{\hat{{X_i}}\hat{{Z_i}}}{2a\hat{Y_i}\hat{U_i}}\cdot\frac{B_{2,i}^3}{3} \cdot\ _2F_1\bigg(2,3;4;-\frac{\hat{V_i}}{\hat{U_i}}B_{2,i}\bigg)
+\frac{\hat{{X_i}}\hat{{Z_i}}\hat{V_i}}{2a\hat{Y_i}\hat{U_i}^2}\cdot\frac{B_{2,i}^2}{2} \cdot\ _2F_1\bigg(2,2;3;-\frac{\hat{V_i}}{\hat{U_i}}B_{2,i}\bigg)\bigg)\nonumber\\
&\quad-\bigg(-\frac{\hat{{X_i}}\hat{{Z_i}}}{2a\hat{Y_i}\hat{U_i}}\cdot\frac{B_{3,i}^3}{3} \cdot\ _2F_1\bigg(2,3;4;-\frac{\hat{V_i}}{\hat{U_i}}B_{3,i}\bigg)
+\frac{\hat{{X_i}}\hat{{Z_i}}\hat{V_i}}{2a\hat{Y_i}\hat{U_i}^2}\cdot\frac{B_{3,i}^2}{2} \cdot\ _2F_1\bigg(2,2;3;-\frac{\hat{V_i}}{\hat{U_i}}B_{3,i}\bigg)\bigg).
\end{align}
Finally, $D'_{23,i}$ in \eqref{e55} is obtained as
\begin{align}\label{e58}
D'_{23,i}
&=
\bigg[\bigg(-\frac{\hat{{X_i}}{\beta_{2,i}}^2}{8a_ib_i\hat{Y_i}\hat{U_i}}\cdot\frac{B_{4,i}^3}{3} \cdot\ _2F_1\bigg(2,3;4;-\frac{\hat{V_i}}{\hat{U_i}}B_{4,i}\bigg)
+\frac{\hat{{X_i}}{\beta_{2,i}}^2\hat{V_i}}{8a_ib_i\hat{Y_i}\hat{U_i}^2}\cdot\frac{B_{4,i}^2}{2} \cdot\ _2F_1\bigg(2,2;3;-\frac{\hat{V_i}}{\hat{U_i}}B_{4,i}\bigg)\bigg)\nonumber\\
&\quad-\bigg(-\frac{\hat{{X_i}}{\beta_{2,i}}^2}{8a_ib_i\hat{Y_i}\hat{U_i}}\cdot\frac{B_{2,i}^3}{3} \cdot\ _2F_1\bigg(2,3;4;-\frac{\hat{V_i}}{\hat{U_i}}B_{2,i}\bigg)
+\frac{\hat{{X_i}}{\beta_{2,i}}^2\hat{V_i}}{8a_ib_i\hat{Y_i}\hat{U_i}^2}\cdot\frac{B_{2,i}^2}{2} \cdot\ _2F_1\bigg(2,2;3;-\frac{\hat{V_i}}{\hat{U_i}}B_{2,i}\bigg)\bigg)\bigg]\nonumber\\
&\quad-\frac{\hat{{X_i}}{\alpha_{1,i}}^2}{8a_ib_i\hat{Y_i}}\bigg(-\frac{\hat{U_i}\tilde{\theta}^3_{L,i}}{3}+
\frac{\hat{V_i}\tilde{\theta}^2_{L,i}}{2}\bigg)\bigg|_{B_{2,i}}^{B_{4,i}}.
\end{align}
  \end{appendices}

\bibliographystyle{IEEEtran}
% argument is your BibTe{X_i} string definitions and bibliography database(s)
\bibliography{myre}

\end{document}